\documentclass[%
 reprint,
nofootinbib,
 amsmath,amssymb,
 aps,
 prc,
]{revtex4-2}

\usepackage{graphicx}% Include figure files
\usepackage{dcolumn}% Align table columns on decimal point
\usepackage{bm}% bold math
\usepackage{romannum} % roman number
\usepackage{subcaption} % This package provides the \subcaptionbox command

\newcommand{\MyRomanNum}[1]{\text{\Romannum{#1}}}
\usepackage{hyperref}% add hypertext capabilities
\usepackage{tikz}
\usetikzlibrary{decorations.pathreplacing,decorations.markings}
\usetikzlibrary{shapes.geometric,positioning}
\tikzset{
operator/.style = {shape = circle, thick, draw, fill, minimum size = 1.2ex,inner sep=0,outer sep=0},
altoperator/.style = {shape = rectangle, thick, draw, fill, minimum size = 1.2ex,inner sep=0,outer sep=0},
fermion/.style = {thick, postaction={decorate}, decoration={markings,mark=at position 0.6 with {\arrow{latex}}}},
undirectedfermion/.style = {thick},
boson/.style = {very thick, dashed},
gaugeboson/.style = {very thick, decorate,decoration={snake}}
}

\newcommand{\sixj}[6]{\begingroup\setlength{\arraycolsep}{0.2em}\begin{Bmatrix} #1 & #2 & #3 \\ #4 & #5 & #6 \end{Bmatrix}\endgroup}

\begin{document}

%\preprint{APS/123-QED}

\title{Factorized approximation to the IMSRG(3)}% Force line breaks with \\

\author{B.C.~He}
\author{S.R.~Stroberg}%
 \email{rstroberg@nd.edu}
\affiliation{%
 Department of Physics and Astronomy,
 University of Notre Dame, Notre Dame IN, 46556 USA
}%

\date{\today}% It is always \today, today,

\begin{abstract}
We describe an approximation to the in-medium similarity renormalization group (IMSRG) method in which we include the effects of intermediate three-body operators arising within nested commutators.
As an initial step, we present the relevant equations for two nested commutators, all of which can be factorized so that the method scales like the standard IMSRG(2) approximation, enabling large-scale calculations.
We test the accuracy of this approximation scheme, and apply it to the isotopic chains of carbon, sulfur and nickel isotopic chains.
We obtain an improved description of spectroscopy, and a reduced dependence on the choice of the valence space.
In addition, we provide an explanation of the relative importance of the diagram topologies included, with an eye toward assessing the impact of remaining omitted terms.
\end{abstract}

\maketitle

\section{Introduction}
In the realm of nuclear theory, accurately solving the many-body Schr\"dinger equation, particularly in predicting properties of atomic nuclei that extend to unexplored areas, remains a formidable challenge. 
Ab initio many-body approaches are designed to offer systematically improvable methods for tackling the many-body problem, in which the nucleon-nucleon two- and three-body nuclear force are adopted as inputs. 
The remarkable advancement in the capabilities of ab initio many-body methods over the past few  decades
can be attributed to both developments in nuclear interactions derived from chiral effective field theory~\cite{Machleidt2011,Epelbaum2009,Hammer2020} and refinements in many-body approaches~\cite{Hergert2020}.
Methods which scale polynomially with system size, particularly the coupled cluster~\cite{Hagen2014}, self-consistent Green's function~\cite{Soma2020}, and in-medium similarity renormalizatinon group (IMSRG)~\cite{Hergert2016} methods have enabled calculations of medium-mass and even heavy nuclei~\cite{Morris2018,Arthuis2020,Miyagi2022,Hu2022,Hebeler2023}.

An essential feature of ab initio calculations is the possibility to provide quantified theoretical uncertainties.
Considerable progress has been made towards quantifying uncertainties related to various aspects of effective field theory~\cite{LENPIC2016,Melendez2017,Melendez2019,Drischler2020,Wesolowski2021,BAND2021,Hu2022}, and truncation of the single-particle basis~\cite{More2013,Furnstahl2014}, but assessing the truncation errors of polynomial-scaling many-body methods requires further understanding.

In this work, we will focus on improving the approximations made within the IMSRG approach.
This framework is especially versatile; it can be applied to ground states of closed- or open-shell nuclei in the single-reference~\cite{Hergert2013} or multi-reference~\cite{Hergert2014} formulations, to excited states within the equations-of-motion framework~\cite{Parzuchowski2017}, to derive effective interactions for direct diagonalization in a valence space~\cite{Stroberg2019}, or to improve convergence properties of the generator coordinate method~\cite{Yao2018} or no-core shell model~\cite{Gebrerufael2017}.
For this reason, understanding and improving the IMSRG truncation scheme will impact a broad range of applications in ab initio nuclear structure.

The essential approximation of the IMSRG is to truncate the operator basis (more details are provided in section~\ref{sec:imsrg}).
Typically this is done by discarding $a$-body operators with $a>2$, denoted the IMSRG(2) approximation.
A treatment that fully includes three-body operators, yielding the IMSRG(3) approximation, has also been explored in small spaces \cite{Heinz2021,IMSRG3_ToBePub}.
However, directly applying IMSRG(3) is practically unfeasible due to the computational expense arising from the underlying three-body operator involved commutators. 
Moreover, it is possible that there are features of the small-space or toy model calculations which do not carry over to realistic large-scale calculations.
In this paper, we present a systematic approach to incorporating induced many-body operators by explicitly evaluating nested commutators, avoiding the explicit construction of intermediate three-body operators.
As a first step, we treat the contribution of intermediate three-body operators to two nested commutators. 

This paper is organized as follows: Section II provides a brief review of the IMSRG. Section III introduces the involved double-commutator and the factorization scheme which enables large-scale calculations. Section IV presents benchmarks against an exactly solvable model as well as against the full IMSRG(3). In section V we apply the method to large-scale calculations of carbon, sulfur, and nickel isotopic chains.

\section{Overview of the IN-MEDIUM SIMILARITY RENORMALIZATION GROUP\label{sec:imsrg}}
The basic approach of the SRG is to perform a unitary transformation on the Hamiltonian
\begin{align}
H(s)\equiv U(s)HU^\dagger(s)
\end{align}
where the unitary transformation $U(s)$ depends on the flow parameter $s$.
The dependence on $s$ is specified by the anti-hermitian generator $\eta$
\begin{equation}
\frac{d }{ds}   U(s) = \eta(s) U(s).
\end{equation}
 
By using the above two equation, we can obtain the SRG flow equation 
\begin{equation}\label{eq_Hflow}
\frac{d }{d s}   H(s) = [ \eta(s),  H(s) ].
\end{equation}
We partition the Hamiltonian into ``diagonal'' and ``off-diagonal'' parts, $H=H^{\rm d}+H^{\rm od}$, and choose a generator such that the off-diagonal part is suppressed $H^{\rm od}(s)\to 0$ as $s\to \infty$.
Throughout this work, we employ the arctangent version of the White generator~\cite{White2002}
\begin{equation}\label{eq_White}
    \langle A | \eta | B \rangle = {\rm atan}\left( \frac{\langle A | H^{\rm od} | B\rangle }{\langle A| H|A\rangle - \langle B|H|B\rangle }\right) 
\end{equation}
where $|A\rangle,|B\rangle$ represent generic $a$-body states with $a=1,2,3,\ldots$
However, the formal developments presented here do not depend on this choice.

While the differential equation~\eqref{eq_Hflow} can be solved directly, it is numerically easier to employ the Magnus formulation of the SRG~\cite{Morris2015},
in which the unitary transformation is expressed as an exponentiated anti-hermitian operator $\Omega(s) = - \Omega^\dagger(s)$
\begin{equation}
U(s) = e^{\Omega(s)}.
\end{equation}

The Baker-Campbell-Hausdorff (BCH) formula is employed to derive a flow equation for the Magnus operator $\Omega(s)$
\begin{align}  \label{Eq_BCH_formula}
\frac{d}{d s} \Omega(s)=\sum_{k=0}^{\infty} \frac{B_{k}}{k !}[\Omega(s),~\eta(s)]^{(k)}
\end{align}
where $B_k$ is the $k$th Bernoulli number and $[\Omega(s),~\eta(s)]^{(k)}$ represent a $k$-fold nested commutator
\begin{equation} \label{Eq_nestedCommutator}
[X,~Y]^{(k)} \equiv
\begin{cases}
\bigl[X, ~[X, ~Y]^{(k-1)}\bigr], & k>0 \\
Y, & k = 0.
\end{cases}
\end{equation}

The flowing Hamiltonian is then given by
\begin{equation} \label{Eq_BCH_transformation}
H(s)= e^{\Omega(s)} H(0) e^{-\Omega(s)} 
= \sum_{k=0}^{\infty} \frac{1}{k !}[\Omega(s), H(0)]^{(k)} 
\end{equation}
where $H(0)$ is the normal-ordered Hamiltonian at $s=0$.
Although~\eqref{Eq_BCH_formula} and~\eqref{Eq_BCH_transformation} formally involve an infinite number of terms, the sum usually converges rapidly and only the first few terms are important.
In practice the nested commutators are evaluated iteratively until the norm of the term is below a given numerical threshold. 

In some cases, the norm of Magnus operator $|| \Omega ||$ grows sufficiently large that is neccesary to evaluate nested commutators out to $k\gtrsim 10$ at each point in the integration of $s$, which makes the calculation costly.
It becomes advantageous to split up the transformation operator $U(s)$
\begin{equation}
U(s) = e^{\Omega(s_n-s_{n-1})} \ldots e^{\Omega(s_2-s_1)} e^{\Omega(s_1)}    
\end{equation}
so that \eqref{Eq_BCH_transformation} becomes
\begin{equation}
    H(s) = e^{\Omega(s_n-s_{n-1})} H(s_{n-1})e^{-\Omega(s_n-s_{n-1})}
\end{equation}
which converges after much fewer nested commutators.

We can also introduce another scheme, which we denote ``hunter-gatherer'', in which we employ only two $\Omega$s~\cite{IMSRG3_ToBePub}.
We designate one operator $\Omega_H$ for collecting information, referred to as the ``hunter'', and the other $\Omega_G$ for storing all transformation information, named the ``gatherer'',
so that 
\begin{equation}
U(s) = e^{\Omega_H(s)} e^{\Omega_G(s)}.
\label{eq_hunter}
\end{equation}
In this scheme, it is also necessary to set a threshold for the size of the hunter.
When the hunter's size exceeds the threshold, we absorb the hunter into the gatherer using the BCH formula
and employ a new hunter to collect information.
In the usual implementation of the IMSRG(2) or IMSRG(3), these details do not have a significant impact on the final result.
However, as we will show below, within the factorized scheme we present in this work, some care is required when splitting up the unitary transformation.

\begin{figure}[t]
\subcaptionbox{$f^\MyRomanNum{1}$}{
  \begin{tikzpicture}
\coordinate (OUT1) at (0.3,1.4);
\coordinate (IN1) at (0.3,-1.2);
%!!!!!!!!! Operator
\node [altoperator] (Omega1) at (-0.3,0.) {};
\node [altoperator] (Omega2) at (-0.3,0.8) {};
\node [operator] (Gamma) at (0.3,-0.6) {};
%!!!!!!!!! connection
\draw [undirectedfermion] (Gamma) to (Omega1);
\draw [undirectedfermion] (Gamma) to [bend right=10] (Omega2);
\draw [undirectedfermion] (Omega1) to[bend right=30] (Omega2);
\draw [undirectedfermion] (Omega1) to[bend right=0] (Omega2);
\draw [undirectedfermion] (Omega1) to[bend left=30] (Omega2);
%!!!!!!!!! in- and out-going lines  
\draw [undirectedfermion] (Gamma) to (OUT1);
\draw [undirectedfermion] (IN1) to (Gamma);
\end{tikzpicture}
}
\subcaptionbox{$f^\MyRomanNum{2}$}{
\begin{tikzpicture}
\coordinate (OUT1) at (0.3,1.4);
\coordinate (IN1) at (0.3,-1.2);
%!!!!!!!!! Operator
\node [altoperator] (Omega1) at (0.3,0.) {};
\node [altoperator] (Omega2) at (-0.3,0.8) {};
\node [operator] (Gamma) at (-0.3,-0.6) {};
%!!!!!!!!! connection
\draw [undirectedfermion] (Gamma) to (Omega1);
\draw [undirectedfermion] (Omega1) to (Omega2);

\draw [undirectedfermion] (Gamma) to [bend right=30] (Omega2);
\draw [undirectedfermion] (Gamma) to [bend right=0] (Omega2);
\draw [undirectedfermion] (Gamma) to [bend left=30] (Omega2);

%!!!!!!!!! in- and out-going lines  
\draw [undirectedfermion] (Omega1) to (OUT1);
\draw [undirectedfermion] (IN1) to (Omega1);
\end{tikzpicture}
}
\subcaptionbox{$f^\MyRomanNum{3}$}{
\begin{tikzpicture}
\coordinate (OUT1) at (0.3,1.4);
\coordinate (IN1) at (0.3,-1.2);
%!!!!!!!!! Operator
\node [altoperator] (Omega1) at (0.3,0.) {};
\node [altoperator] (Omega2) at (-0.3,0.8) {};
\node [operator] (Gamma) at (0.3,-0.6) {};
%!!!!!!!!! connection
\draw [undirectedfermion] (Gamma) to (Omega1);

\draw [undirectedfermion] (Gamma) to [bend left=20] (Omega2);
\draw [undirectedfermion] (Gamma) to [bend left=40] (Omega2);

\draw [undirectedfermion] (Omega1) to[bend right=10] (Omega2);
\draw [undirectedfermion] (Omega1) to[bend left=10] (Omega2);

%!!!!!!!!! in- and out-going lines  
\draw [undirectedfermion] (Omega1) to (OUT1);
\draw [undirectedfermion] (IN1) to (Gamma);
\end{tikzpicture}
}\\ 
\subcaptionbox{$\Gamma^\MyRomanNum{1}$}{
\begin{tikzpicture}
\coordinate (OUT1) at (-0.6,1.4);
\coordinate (OUT2) at (0.6,1.4);
\coordinate (IN1) at (-0.6,-1.2);
\coordinate (IN2) at (0.6,-1.2);
\node [altoperator] (Omega1) at (-0.2,0.2) {};
\node [altoperator] (Omega2) at (-0.45,1.0) {};
\node [operator] (Gamma) at (0,-0.4) {};
\draw [undirectedfermion] (Gamma) to (Omega1);
\draw [undirectedfermion] (Omega1) to[bend left=40] (Omega2);
\draw [undirectedfermion] (Omega1) to[bend right=40] (Omega2);
\draw [undirectedfermion] (Omega2) to[bend left=0] (Omega1);
\draw [undirectedfermion] (Omega2) to (OUT1);
\draw [undirectedfermion] (Gamma) to (OUT2);
\draw [undirectedfermion] (IN1) to (Gamma);
%\draw [fermion] (IN2) to (Gamma);
\draw [undirectedfermion] (IN2) to (Gamma);
\end{tikzpicture}
}
\subcaptionbox{$\Gamma^\MyRomanNum{2}$}{
\begin{tikzpicture}
\coordinate (OUT1) at (-0.6,1.4);
\coordinate (OUT2) at (0.6,1.4);
\coordinate (IN1) at (-0.6,-1.2);
\coordinate (IN2) at (0.6,-1.2);
!!!!!!!!! Operator
\node [altoperator] (Omega1) at (0.2,0.1) {};
\node [altoperator] (Omega2) at (-0.4,0.8) {};
\node [operator] (Gamma) at (-0.1,-0.4) {};
!!!!!!!!! connection
\draw [undirectedfermion] (Gamma) to (Omega1);
\draw [undirectedfermion] (Gamma) to [bend right=0] (Omega2);
\draw [undirectedfermion] (Omega2) to[bend right=30] (Gamma);
\draw [undirectedfermion] (Gamma) to [bend right=25] (Omega2);
!!!!!!!!! in- and out-going lines  
\draw [undirectedfermion] (Omega2) to (OUT1);
\draw [undirectedfermion] (Omega1) to (OUT2);
\draw [undirectedfermion] (IN1) to (Gamma);
%\draw [fermion] (IN2) to (Gamma);
\draw [undirectedfermion] (IN2) to (Omega1);
\end{tikzpicture}
}
\subcaptionbox{$\Gamma^\MyRomanNum{3}$}{
\begin{tikzpicture}
\coordinate (OUT1) at (-0.6,1.4);
\coordinate (OUT2) at (0.6,1.4);
\coordinate (IN1) at (-0.6,-1.2);
\coordinate (IN2) at (0.6,-1.2);
\node [altoperator] (Omega1) at (0.,0.2) {};
\node [altoperator] (Omega2) at (-0.4,0.9) {};
\node [operator] (Gamma) at (0,-0.4) {};
\draw [undirectedfermion] (Gamma) to (Omega1);
\draw [undirectedfermion] (Omega1) to[bend left=20] (Omega2);
\draw [undirectedfermion] (Omega1) to[bend right=20] (Omega2);
\draw [undirectedfermion] (Gamma) to[bend left=20] (Omega2);
\draw [undirectedfermion] (Omega2) to (OUT1);
\draw [undirectedfermion] (Omega1) to (OUT2);
\draw [undirectedfermion] (IN1) to (Gamma);
%\draw [fermion] (IN2) to (Gamma);
\draw [undirectedfermion] (IN2) to (Gamma);
\end{tikzpicture}
}
\subcaptionbox{$\Gamma^\MyRomanNum{4}$}{
\begin{tikzpicture}
\coordinate (OUT1) at (-0.6,1.4);
\coordinate (OUT2) at (0.6,1.4);
\coordinate (IN1) at (-0.6,-1.2);
\coordinate (IN2) at (0.6,-1.2);
\node [altoperator] (Omega1) at (-0.0,0.2) {};
\node [altoperator] (Omega2) at (-0.3,0.8) {};
\node [operator] (Gamma) at (-0.2,-0.4) {};
\draw [undirectedfermion] (Gamma) to (Omega1);
\draw [undirectedfermion] (Omega1) to (Omega2);
\draw [undirectedfermion] (Omega2) to[bend right=30] (Gamma);
\draw [undirectedfermion] (Omega2) to[bend left=0] (Gamma);
\draw [undirectedfermion] (Omega2) to (OUT1);
\draw [undirectedfermion] (Omega1) to (OUT2);
\draw [undirectedfermion] (IN1) to (Gamma);
%\draw [fermion] (IN2) to (Gamma);
\draw [undirectedfermion] (IN2) to (Omega1);
\end{tikzpicture}
}
\caption{Hugenholtz skeleton diagrams indicating the topologies of the double commutators are illustrated as follows: the upper three plots, labeled (a) to (c), are topologies for the one-body operators, while the lower four plots are the topologies for the two-body operators, labeled (d) to (g).}
\label{fig_topology}
\end{figure}
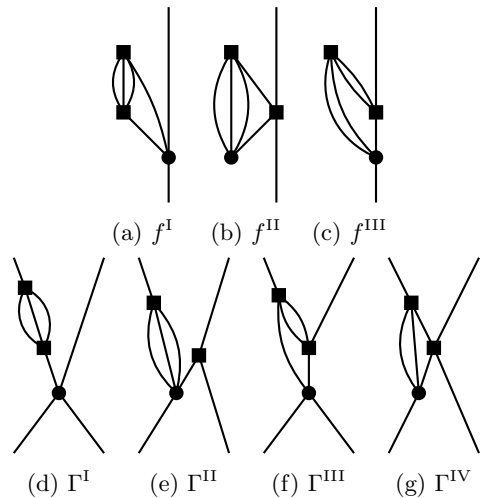

We employ a Fock space representation and normal order all operators with respect to a reference state $|\Phi_0\rangle$, so that the Hamiltonian is written as
\begin{align}
H & = E_0 + \sum_{ij} f_{ij} \{a_i^\dagger a_j\} + \frac{1}{4}\sum_{ijkl} \Gamma_{ijkl} \{a_i^\dagger a_j^\dagger a_l a_k \} \nonumber\\
    &+ \frac{1}{36}\sum_{ijk lmn} W_{ijklmn} \{a_i^\dagger a_j^\dagger a_k^\dagger a_n a_m a_l \}
\end{align}
where we denote the normal ordering with braces.
In the usual IMSRG(2) approximation, all operators are truncated at the two-body level, including intermediate nested commutators appearing in~\eqref{Eq_BCH_transformation}.
For the valence-space calculations presented in this work, we adopt an ensemble reference corresponding to the spherical Hartree-Fock ground state in the equal-filling approximation~\cite{Stroberg2017}.

\section{Factorized double commutators of the IMSRG(3)}
\label{sec:Double_commutators}
The IMSRG method has been extended to a full three-body treatment, indicated IMSRG(3), demonstrating a systematic improvement over IMSRG(2) towards the exact result~\cite{Heinz2021}. 
The commutators involved in IMSRG(3) are more computationally expensive, with a complexity ranging from $N^7$ to $N^9$.
This is significantly worse than IMSRG(2), where the worst-case complexity is $n^6$, and renders large-scale calculations infeasible.

In this work, we explore a new approximation to IMSRG, wherein we can avoid the direct construction of higher-body operators.
The key point is that, for a reference state $\Phi_0$ that sufficiently approximates the exact wave function, the expectation values of many-body operators are suppressed by $N_{\rm qp}/A$ where $N_{qp}$ is the number of quasi-particles (particles plus holes with respect to $\Phi_0$)~\cite{Friman2011}.
However, during the IMSRG flow, or within nested commutators, induced many-body operators can subsequently contract down to one- and two- body operators, and this effect is not suppressed.

More concretely, the commutator of two two-body operators, e.g. $\Omega$ and $H$, will include a three-body part, which we denote $[\Omega,H]_{\rm 3b}$.
This induced three-body operator can then be contracted back down to  one-body and two-body pieces in a double commutator.
We can incorporate this effect by adding a correction $\Delta_{XY}^{(k)}$ to \eqref{Eq_nestedCommutator}, defined by
\begin{equation}
\Delta_{XY}^{(k)} =
\begin{cases}
[X,[X,[X,Y]_{\rm 2b}^{(k-2)}]_{\rm 3b}]_{\rm 1b,2b}, & k>1 \\
0, & k\leq 1.
\end{cases}
\end{equation}
As a specific example, taking $X=\Omega$, $H=Y$, and $k=2$ we have
\begin{equation} \label{eq_induced1b2b}
\Delta_{\Omega H}^{(2)} = [ \Omega_{\rm 2b}~, [\Omega_{\rm 2b}~, H_{\rm 2b} ]_{\rm 3b} ]_{\rm 1b,2b}.
\end{equation}
This type of correction has been explored in the context of quantum chemistry~\cite{Neuscamman2009}, and in shell model coupled cluster (SMCC)~\cite{Sun2018}.
Importantly, while the double commutator in \eqref{eq_induced1b2b} naively scales as $N^7$, it can be factorized~\cite{Morris2016,Neuscamman2009,Sun2018} to yield expressions that scale as $N^6$, comparable to the scaling of IMSRG(2).

Note that by modifying the expression~\eqref{Eq_nestedCommutator}, we do not just include corrections to the transformed operator in~\eqref{Eq_BCH_transformation} at two nested commutators.
We also include corrections to arbitrary nested commutators in which induced three-body operators are immediately contracted back down to one- or two-body operators.
For example, $[\Omega,H]^{(4)}$ would receive a contribution
\begin{equation}
    [\Omega,[\Omega,[\Omega,[\Omega,H]_{\rm 3b}]_{\rm 2b}]_{\rm 3b}]_{\rm 2b}.
\end{equation}
It is straightforward to extend this type of correction so that triple-nested commutators are evaluated exactly, but we leave that for future work, noting that not all of these diagrams can be factorized to scale as $N^6$.

In the following, we will use notation which assumes $X=\Omega$ and $Y=H$, but the expressions can be directly applied to other operators (with modified angular momentum coupling for operators that are not scalars under rotation).
The one-body part of the correction \eqref{eq_induced1b2b} consists of three distinct diagram topologies, illustrated in Fig.~\ref{fig_topology}(a)-(c)
\begin{equation}
\Delta^{(2)}_{\Omega H,{\rm 1b}} =  f^\MyRomanNum{1}  +  f^{\MyRomanNum{2}}  + f^{\MyRomanNum{3}}  .
\end{equation} 
We provide an explicit expression for only one example here; the others can be found in appendix \ref{appendix:M-scheme}.
The one-body matrix elements of $f^\MyRomanNum{1}$ are given by
\begin{align} \label{eq_DoubleCommutator_onebody}
 f^\MyRomanNum{1}_{ij}  =&\frac{1}{2} \sum_{abcde} \left( \bar{n}_a \bar{n}_b  {n}_c  {n}_d  -  {n}_a  {n}_b \bar{n}_c  \bar{n}_d  \right)   \nonumber \\
  &\times \left( \Omega_{cdab}\Omega_{abce}   \Gamma_{eidj}  +  \Omega_{cdab} \Omega_{abce}  \Gamma_{diej}  \right)
\end{align}
where $n_a$ is the occupation of orbit a, and $\bar n_a = 1 - n_a$. 
(Since this is a double commutator, there are four orderings of the operators involved in the expression.)

The two-body part of the correction~\eqref{eq_induced1b2b} consist of four topologies, illustrated in Fig.~\ref{fig_topology}(d)-(g):
\begin{align}
\Delta^{(2)}_{\Omega H,{\rm 2b}} =& \Gamma^{\MyRomanNum{1}} + \Gamma^{\MyRomanNum{2}} + \Gamma^{\MyRomanNum{3}}  +\Gamma^{\MyRomanNum{4}} .
\end{align}
Again, to illustrate we give the expression for one of the topologies.
The others can be found in Eq. (\ref{DoubleCommutator_onebody_all}) in the appendix \ref{appendix:M-scheme}.
\begin{align} \label{eq_DoubleCommutator_twobody}
\Gamma^{\MyRomanNum{1}}_{ijkl}  =&\frac{1}{2} \sum_{abcd} \left( \bar{n}_a  \bar{n}_b {n}_c  +  {n}_a  {n}_b  \bar{n}_c \right) \nonumber\\
& \times \left\{ \left(1 - \hat P_{ij} \right) \Omega_{ciab} \Omega_{abcd} \Gamma_{djkl} \right. \nonumber\\
& + \left.  \left(1 - \hat P_{kl} \right) \Omega_{adcb} \Omega_{cbak} \Gamma_{ijdl} \right\}
\end{align}
where the permutation operator $\hat P_{ij}$ exchanges the indices $i$ and $j$ of the
expression to its right. 
Although the double commutators avoid the explicit construction of three-body operators, the complexity is the same as that of IMSRG(3). While this reduces memory requirements, it does not improve computational speed.

As mentioned above, the expressions can be factorized to reduce the scaling to $N^6$.
To illustrate, we take the expression in \eqref{eq_DoubleCommutator_onebody}, rewrite it as
\begin{equation}\label{eq_doubleCommutator_1b_factorized}
 f^\MyRomanNum{1}_{ij}  = \sum_{ab} \left( \chi^{\alpha}_{ab}  \Gamma_{biaj} + \chi^{\alpha}_{ab} \Gamma_{aibj} \right) 
\end{equation}
where we have introduced the intermediate object
\begin{align} \label{eq_chi(1b)_a}
\chi^{\alpha}_{ij} =\frac{1}{2} \sum_{abc} \left( \bar{n}_a \bar{n}_b  {n}_c  {n}_i  -  {n}_a {n}_b \bar{n}_c  \bar{n}_i \right)  \Omega_{ciab} \Omega_{abcj} 
\end{align}
Equation~\eqref{eq_doubleCommutator_1b_factorized} scales as $N^4$, while \eqref{eq_chi(1b)_a} scales as $N^5$, so the original $N^7$ scaling of \eqref{eq_DoubleCommutator_onebody} is reduced to $N^4+N^5$.

\begin{figure}[t]
    \centering
        \includegraphics[width=0.97\linewidth]{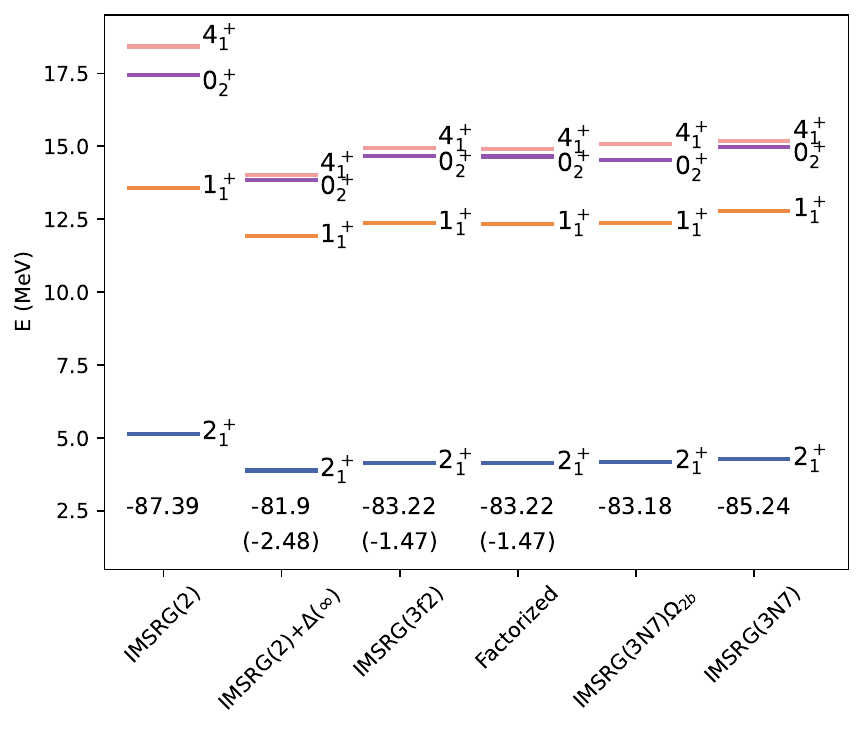}
        \caption{Excited state spectrum of $^{12}$C computed with different levels of approximation from IMSRG(2) to IMSRG(3). The ground state energy without the triples correction is indicated below the spectra, with the triples correction in parentheses. See text for more details.
    %\BCH{Didn't get result from IMSRG3 yet.} 
    \label{fig_C12spectrum}
    }
\end{figure}

In order to take advantage of rotational invariance, we employ $J$-coupled operators.
The corresponding $J$-scheme factorized equations are presented in the appendix~\ref{section::Jscheme_factorizedDCs}.

We indicate this approximation  IMSRG(3f$_2$).
The ``3'' indicates the inclusion of intermediate three-body operators, ``f'' signifies the use of a factorized method to evaluate nested commutators, and the subscript ``2'' denotes that only double commutators are factorized in this work.
This naming convention is designed to be extendable to future extensions, wherein variations such as IMSRG(3f$_3$) and IMSRG(4f$_3$) may be employed.
We note that this scheme depends on the Magnus formulation; implementing this within the flow equation formulation is less straightforward.

Before presenting our results, we wish to discuss further the connection between the IMSRG(3f$_2$) approximation and the SMCC.
While the two methods are very similar, there are a few important differences.
First the SMCC is formulated as a similarity transformation, so that the resulting effective interaction is non-Hermitian, though it can be rendered Hermitian by an additional step~\cite{Sun2021};
the IMSRG produces Hermitian interactions and operators which may be directly used with standard shell model codes.

Second, in ref.~\cite{Sun2018}, the normal ordering was performed with respect to the core of the valence space, while here we employ an ensemble reference corresponding to the nucleus of interest, such that some valence orbits have non-zero occupations.
The occupied valence orbits approximately capture the induced many-body forces between valence nucleons.
When the normal ordering is performed with respect to the core, the importance of explicit valence three-body forces will grow rapidly with the number of valence particles.
In ref.~\cite{Sun2018}, it was found that explicitly including three-body forces in the valence-space diagonalization had a non-negligible impact.
(We note that the VS-IMSRG results presented in ref.~\cite{Sun2018} use the core as the reference).
In ref.~\cite{Sun2021}, a particle-hole framework analogous to the ensemble reference was adopted, though in that case only a subset of the nested commutator expressions were retained.

Finally, in the SMCC it appears that the three-body cluster operator, denoted $S_{\rm 3b}$ has a non-negligible impact.
In the IMSRG, the analogous contribution would be the three-body part of $\Omega$ (or $\eta$).
The perturbative triples correction we include captures the leading part of this term, but we have not included the one- and two-body valence operators induced by $\Omega_{\rm 3b}$.
Such terms \emph{are} included in the IMSRG(3) and IMSRG(3N7) approximations which we use in the benchmarks in section~\ref{sec:benchmarks}, and the comparison indicates that their effect is very small.

\section{Benchmarking IMSRG(\texorpdfstring{${\rm 3f}_2$}{3f2})\label{sec:benchmarks}}
In this section, we will examine factorized double commutators and IMSRG(3f$_2$). We aim to demonstrate the effectiveness and efficiency of factorized double commutators and explain how to use them in practical calculations.
Unless otherwise noted, for all calculations in this paper we use the 1.8/2.0 (EM) NN+3N interaction~\cite{Hebeler2011} in an oscillator basis with $\hbar\omega=16$~MeV.

\subsection{Dependence on solution strategy}

While the factorized corrections described in section~\ref{sec:Double_commutators} nominally scale as $N^6$, some of the terms are still significantly more costly than the IMSRG(2) commutators.
Specifically, the topologies $\Gamma^{\MyRomanNum{3}}$ and $\Gamma^{\MyRomanNum{4}}$, which involve a two-body intermediate $\chi$, involve multiple angular momentum recouplings.
While these terms have a non-negligible effect, we have found that they terms have a relatively minor impact on the generator $\eta$ during the flow (see further discussion below).
Consequently it is an excellent approximation to neglect them during the flow and only include them at the end when evaluating~\eqref{Eq_BCH_transformation} with $\Omega(\infty)$.

\begin{figure}[t]
    \centering
    \includegraphics[width=0.97\linewidth]{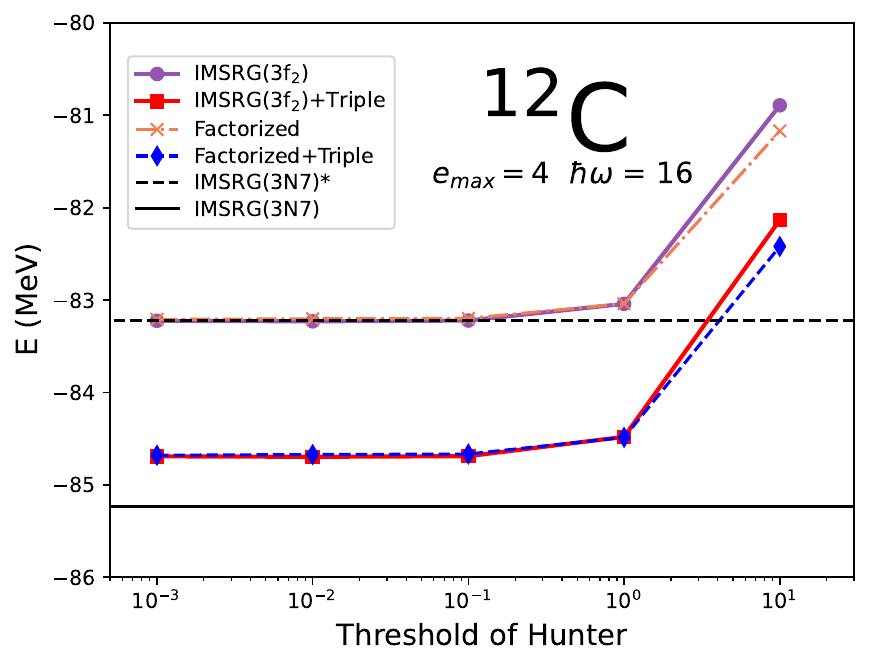}
    \caption{Ground state energy of $^{12}$C computed in the VS-IMSRG for several approximation schemes (indicated in the legend) as a function of the threshold parameter for $\Omega_{H}$ used in the hunter-gatherer approach (see text for details). In the legend ``Triple'' indicates the inclusion of perturbative triples.
    }
    \label{fig_benchmark_C12}
\end{figure}

To illustrate the impact of the various corrections that enter when going from IMSRG(2) to IMSRG(3), we compute the low-lying excitation spectrum of $^{12}$C in a single-particle space defined by  $e_{\rm max}=4$, and a $p$-shell valence space.
As a first correction, we perform an IMSRG(2) decoupling to obtain $\Omega$, and then transform the Hamiltonian via \eqref{Eq_BCH_transformation} using the correction \eqref{Eq_nestedCommutator}.
We label this IMSRG(2)+$\Delta(\infty)$, because we only apply the correction at $s=\infty$.
Next, we use the IMSRG(3f$_2$) approximation, in which factorized corrections involving one-body intermediates are included during the IMSRG flow, and the corrections involving two-body intermediates are included at $s=\infty$.
We then include all factorized corrections during the flow, indicated ``factorized'' in Figs.~\ref{fig_C12spectrum} and~\ref{fig_benchmark_C12}; this approximation is equivalent  to retaining a subset of the IMSRG(3) commutators, namely\footnote{Here we use the notation $[a,b]_c$ to indicate the $c$-body part of the commutator of an $a$-body operator with a $b$-body operator. The scaling of such a term is $N^{a+b+c}$.} $[2,2]_3$, $[2,3]_2$, and $[2,3]_1$, which we denote IMSRG(3N7)$^*$ and neglecting the three-body part of $\Omega$.
Next, we include all IMSRG(3) commutators which scale as $N^7$ or better, corresponding to IMSRG(3N7)$^*$ plus $[3,3]_0$, $[3,3]_1$, $[1,3]_3$, and $[1,3]_2$; we indicate this IMSRG(3N7).
We may also use the IMSRG(3N7) approximation without a three-body part of $\Omega$, which we indicate IMSRG(3N7)$\Omega_{\rm 2b}$.
%Finally, we include all IMSRG(3) commutators througout the flow, indicated IMSRG(3).
All of these are presented in Fig.~\ref{fig_C12spectrum}.

The ground state energies are shown at bottom of each column. In IMSRG(2)+$\Delta(\infty)$, IMSRG(3f$_2$) and the factorized approximation, the pertuibetive triples are included for the ground states.

The biggest change comes from simply including the factorized commutator terms at $s=\infty$, without modifying the IMSRG flow.
A smaller, but still non-negligible shift comes from including the factorized commutators with one-body intermediates.
Beyond this, the corrections to the spectrum are minor and, in our assessment, generally not worth the considerable additional computational cost.
Including the factorized terms at $s=\infty$ shifts the ground state energy by +5.5~MeV, while the perturbative triples contribute -2.5~MeV.
Including the factorized commutators involving one-body intermediates in the IMSRG(3f$_2$) approximation almost exactly recovers the full IMSRG(3N7) result with a two-body $\Omega$.
The IMSRG(3N7) approximation, including a three-body $\Omega$ slightly modifies the spectrum and adds an additional -2.06~MeV to the ground state energy, -1.47 of which is captured by the perturbative triples correction.

In addition, when employing the hunter-gatherer scheme~\eqref{eq_hunter}, we must choose the threshold on $\|\Omega_{H}\|$ at which we gather $\Omega_{H}$ into $\Omega_{G}$, and ideally the results should not depend on this arbitrary parameter.
In Fig.~\ref{fig_benchmark_C12}, we compute the ground state energy of $^{12}$C using the same interaction and model space as for Fig.~\ref{fig_C12spectrum}.
The results are shown as a function of the threshold parameter on $\|\Omega_H\|$.
We find that the final energies are threshold-independent below a threshold of about 1.
For larger values of the threshold, $\Omega_{G}$ is of comparable size to $\Omega_H$ and the cross terms like $[\Omega_G,[\Omega_H,H]_{\rm 3b}]]$, which are missed by the IMSRG($3f_2$) scheme, become non-negligible.

 For the IMSRG(3), IMSRG(3N7), and IMSRG(3N7)$^*$ approximations, we do not explicitly include the induced three-body forces in the valence space diagonalization.
 Instead, after the IMSRG decoupling, we discard the residual three-body force before re-normal ordering with respect to the core.

To illustrate the efficiency of the IMSRG(3f$_2$) approximation, we compare the CPU time required for a valence space IMSRG decoupling for $^{12}$C within the model space e$_{max}$ = 4,
utilizing a single node with 64 cores at the Center for Research Computing at Notre Dame.
For IMSRG(3N7) and the full IMSRG(3), the CPU times were 8.5 hours and 94 hours, respectively.
In contrast, IMSRG(3f$_2$) required only 15 seconds, comparable with the 12 seconds required for IMSRG(2).

\subsection{Benchmarking with an exactly-solvable model}

\begin{figure*}
    \centering
    \includegraphics[width=\textwidth]{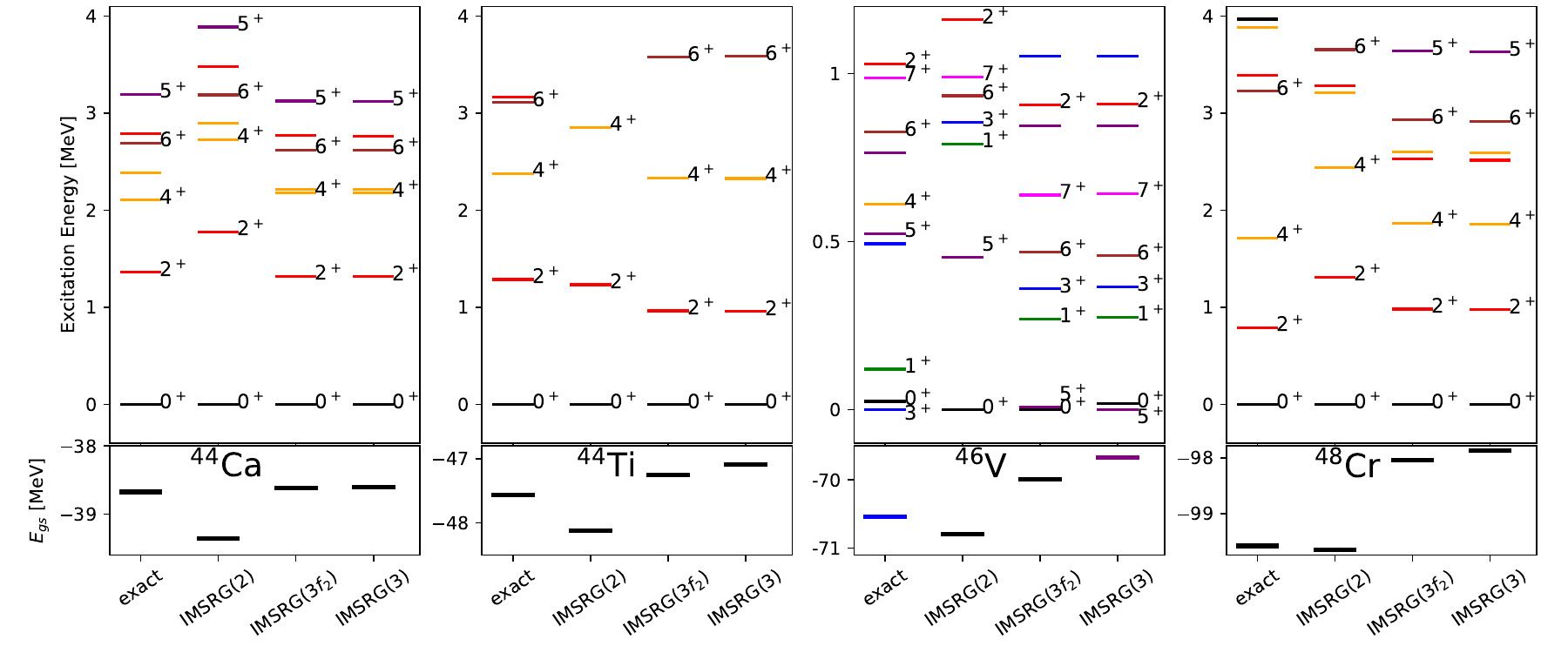}
    \caption{\label{fig_gx1a}Excitation spectra and ground state energies using the gxpf1a interaction~\cite{Honma2005} in the $pf$ shell. For the IMSRG calculations, we decouple the $f_{7/2}$ orbits from $\{f_{5/2},p_{3/2},p_{1/2}\}$  and diagonalize in the reduced space.}
\end{figure*}

Before moving to large-scale calculations, we benchmark the IMSRG($3f_2$) approximation in a tractable model space so that we may compare with the exact solution and the full IMSRG(3) result.
For this purpose, we consider the empirical gxpf1a interaction in the $fp$ shell~\cite{Honma2005}.
We use the VS-IMSRG to decouple the $f_{7/2}$ orbits from $\{p_{3/2}, p_{1/2},f_{5/2}\}$, and then diagonalize in the reduced valence space.
It is worth noting that in this simplified model there are no core orbits, so there may be contributions that are important in large-scale calculations which are identically zero here.
Fig.~\ref{fig_gx1a} shows the exact spectra and ground state energies compared with three levels of approximation: IMSRG(2), IMSRG($3f_2$), and IMSRG(3).
As with the IMSRG(3N7) approximation in the previous section, for the IMSRG(3) calculation we do not include the residual induced three-body force in the valence space diagonalization.
The first striking observation is that the IMSRG($3f_2$) spectra are remarkably close to the full IMSRG(3).
In addition, the IMSRG(3) results are often, but not always, closer to the exact value than the IMSRG(2) results.

There are two potential sources of the remaining error in the IMSRG(3) approximation.
The first arises from terms involving $a$-body operators with $a\geq 4$, and the second is due to the neglect of induced valence three-body operators in the final diagonalization.
In light of the findings in the context of SMCC~\cite{Sun2018}, we suspect that the valence three-body operators may become important at this level of precision.
We will explore this in a future study.

\subsection{Relative importance of topologies\label{sec:topologies}}
\begin{figure}[ht]
    \centering
    \includegraphics[width=\linewidth]{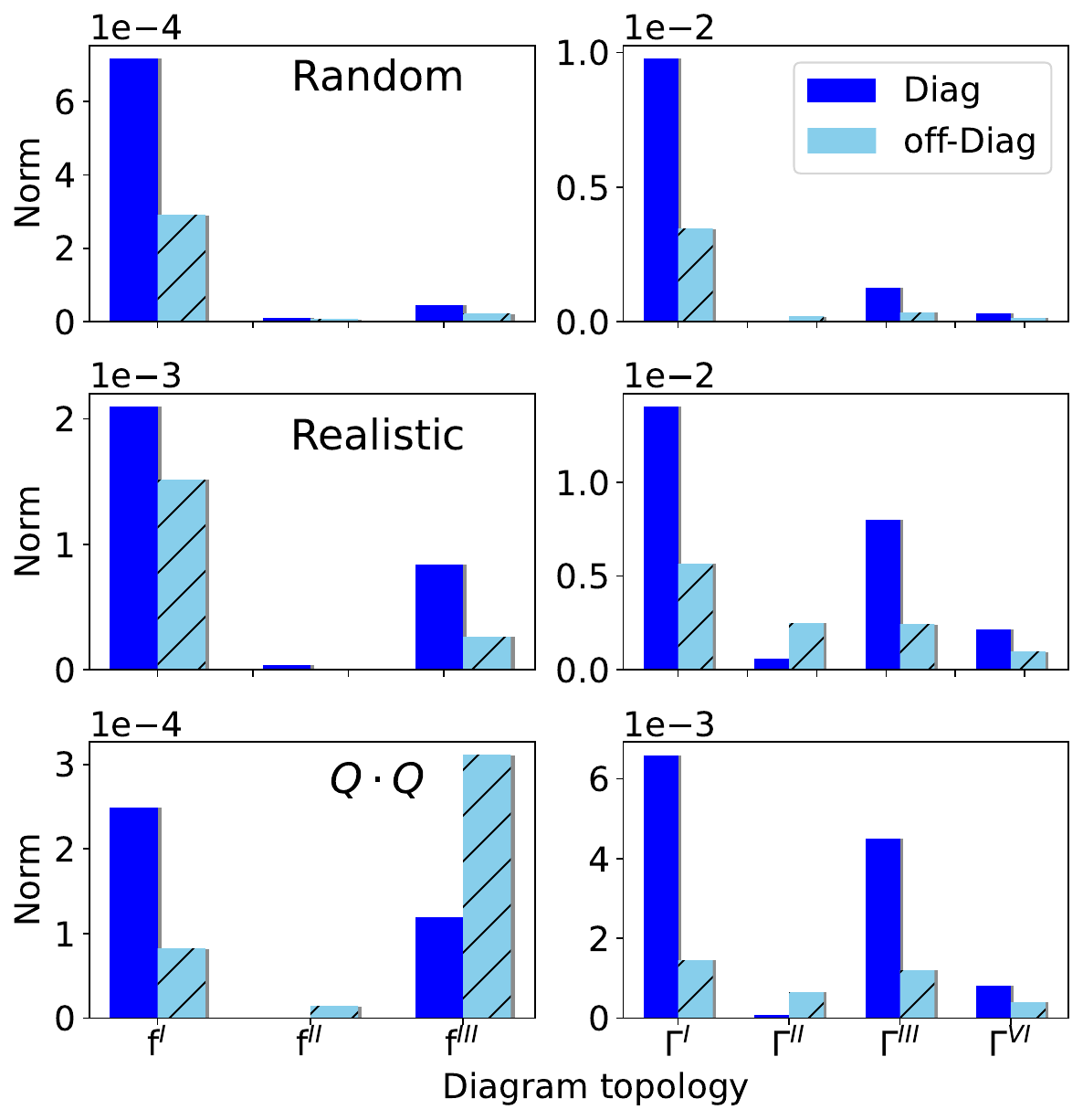}
    \caption{The norm of the one- and two-body operators generated by double commutators, broken down by topology of diagrams (labels correspond to Fig.~\ref{fig_topology}) and whether they are diagonal or off-diagonal.
    These were generated using a random interaction (top row), a realistic interaction (middle row) and a separable $Q\cdot Q$ interaction (bottom row).
    See text for more details.
    }
    \label{fig_NormOfDoubleCommutators}
\end{figure}

Understanding which terms are most important and why may provide insight into how to make further improvements.
To evaluate the significance of these diagrams, we compute the double commutator \eqref{eq_induced1b2b} between two two-body operators $\Omega$ and $H$ with $e_{max}=4$. 
The Frobenius norms of the final one- and two-body operators are displayed in Fig.~\ref{fig_NormOfDoubleCommutators}, where we split the operator into diagonal and off-diagonal pieces, in the sense of equation~\eqref{eq_White}.
The plots in the top row, labeled ``Random'' were generated using $\Omega$ and $H$ operators with random matrix elements drawn from a Gaussian distribution centered on zero.
For the plots in the middle row, $H$ was taken to be the 1.8/2.0 (EM) interaction normal-ordered with respect to a $^{12}$C referece, and $\Omega$ was obtained by decoupling the $p$-shell valence space.
For the bottom row, we take $H=Q\cdot Q$ where $Q_{\mu}=r^2Y_{2\mu}(\theta,\phi)$ is the mass quadrupole operator, and $\Omega$ is taken to be the off-diagonal part of $H$.

The $H$ and $\Omega$ operators are scaled to match the norm of their realistic counterparts, and
all plotted norms have been divided by $\|H \| \cdot \|\Omega\|^2$, where the double bars indicate the Frobenius norm.

For the random operators in the top row of Fig.~\ref{fig_NormOfDoubleCommutators}, the one-body piece of the double commutator is dominated by the $f^{\MyRomanNum{1}}$ topology,
while the two-body piece is dominated by the $\Gamma^{\MyRomanNum{1}}$ topology.
This is also true for the realistic case.
Both of these topologies involve one-body intermediates. 
In the realistic case, the dominance of the $f^{\MyRomanNum{1}}$ and $\Gamma^{\MyRomanNum{1}}$ topologies is less pronounced;
the diagrams $f^{\MyRomanNum{3}}$ and $\Gamma^{\MyRomanNum{3}}$ are sub-leading, but cannot be disregarded.

Comparing the diagonal and off-diagonal pieces of the norm, we find that the topologies involving intermediate one-body operator, specifically  f$^{\MyRomanNum{1}}$, f$^{\MyRomanNum{2}}$, $\Gamma^{\MyRomanNum{1}}$, and $\Gamma^{\MyRomanNum{2}}$, generate the majority of the off-diagonal part of the resulting operator.
This explains why the IMSRG(3f$_2$) approximation is effective.
When solving the SRG flow equation \eqref{eq_Hflow} or \eqref{Eq_BCH_formula}, the generator $\eta$ depends on the off-diagonal piece of the Hamiltonian; the diagonal piece enters only indirectly through subsequent commutators of the form $[\eta,H^{\rm d}]^{\rm od}$. 
Therefore, during the flow, we can disregard other topologies and only include those operators involving one-body intermediate operator.
After obtaining the Magnus operator $\Omega(s=\infty)$ the other terms are included in the transformation~\eqref{Eq_BCH_transformation}.

We would like to understand the relative importance of the topologies as revealed in Fig.~\ref{fig_NormOfDoubleCommutators}.
First, let us consider the topology $\Gamma^{\MyRomanNum{1}}$, which is especially dominant for a random operator, suggesting that the enhancement mechanism is independent of the specific features of the nucleon-nucleon interaction.
Schematically, the factorized expression for $\Gamma^{\MyRomanNum{1}}$ is (see appendix for full expressions)
\begin{equation} \label{eq_GammaI}
    \Gamma^{\MyRomanNum{1}}_{ijkl} \sim \sum_a \chi^{\epsilon}_{ai}\Gamma_{ajkl}
\end{equation}
with the one-body intermediate
\begin{equation}\label{eq_chi_epsilon}
    \chi^{\epsilon}_{ij} \sim \sum_{bcd} \Omega_{ciab}\Omega_{abcj}.
\end{equation}
We focus our attention on the diagonal elements $\chi^{\epsilon}_{ii}$.
In that case, the matrix elements in the sum in \eqref{eq_chi_epsilon} become $\Omega_{ciab}\Omega_{abci}=-|\Omega_{ciab}|^2$.
All terms in the sum manifestly have the same sign, so that the diagonal elements of $\chi^{\epsilon}$ are large and negative.
Considering the effect of these diagonal elements on $\Gamma^{\MyRomanNum{1}}$, we see from \eqref{eq_GammaI} that we have $\Gamma^{\MyRomanNum{1}}_{ijkl} \sim -|\chi^{\epsilon}_{ii}|  \Gamma_{ijkl}$; that is, the correction to $\Gamma_{ijkl}$ is negative and proportional to $\Gamma_{ijkl}$.
We generically obtain suppression of all two-body matrix elements.

\begin{figure}[ht]
    \centering
    \includegraphics[width=\linewidth]{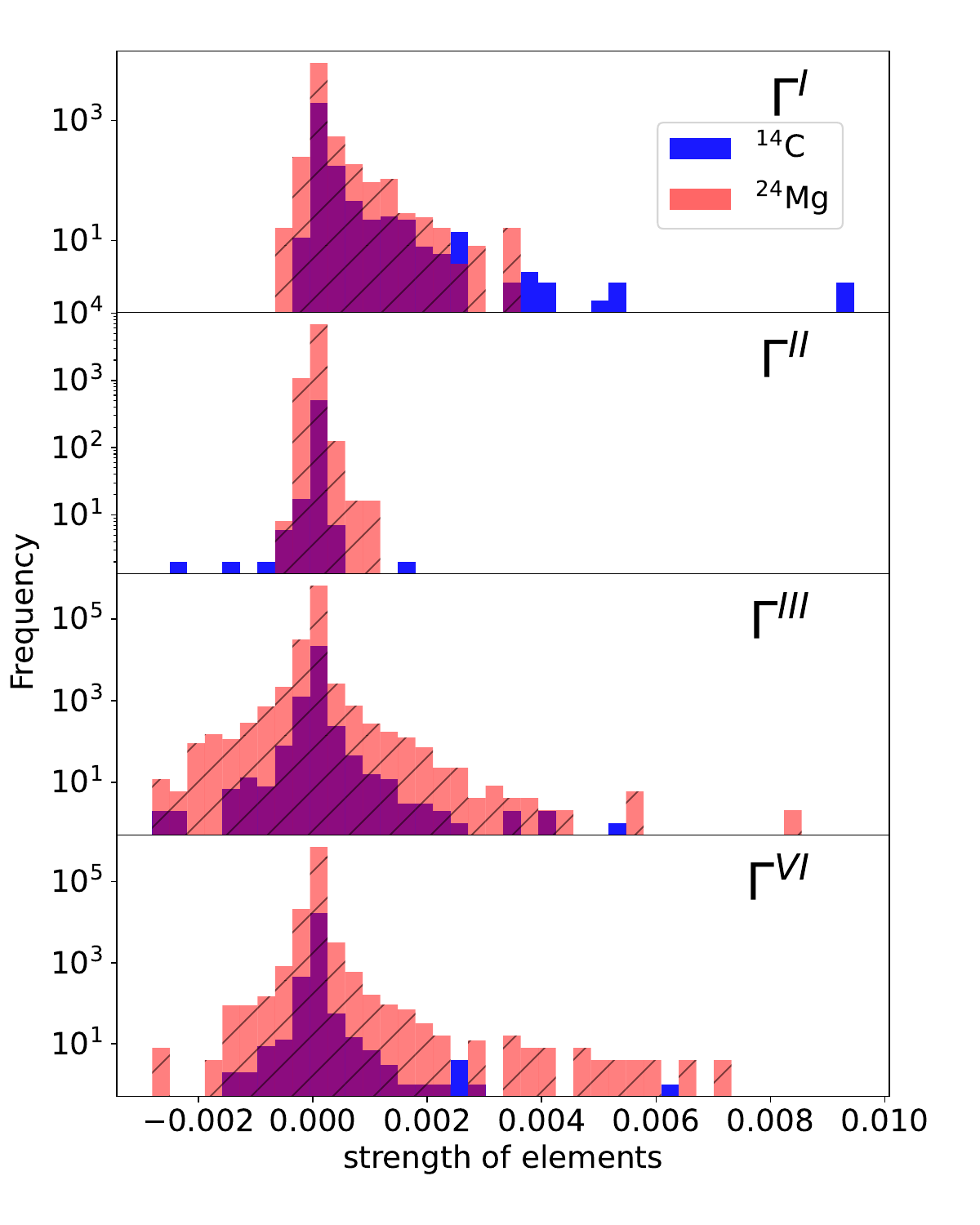}
    \caption{The distribution of terms entering the the sums corresponding to diagrams I, II, III and VI of two body operator. The distribution is computed decoupling the $p$ shell with a $^{14}$C reference (blue) and decoupling the $sd$ shell with a $^{24}$Mg reference (red with hashes). 
    }
    \label{fig_histograms}
\end{figure}

We demonstrate this same-sign coherent enhancement in Fig.~\ref{fig_histograms}, which shows histograms of the individual terms in the sums for each diagram topology.
All the terms in the topology $\Gamma^{\MyRomanNum{1}}$ are positive, while the other topologies have positive and negative terms with essentially equal probability.
To evaluate the double commutators for these histograms, we used the same interaction and single-particle space as before, and normal order with respect
to two reference states, $^{14}$C and $^{24}$Mg, to confirm that the behavior is robust.
The Magnus operator $\Omega$ is obtained from solving VS-IMSRG(2) to decouple the valence space, defined as the $p$-shell above a $^4$He core and $sd$-shell above an $^{16}$O core for $^{14}$C and $^{24}$Mg, respectively.
The distributions show contributions to the $J=0$ proton-proton matrix element $\langle aa J| \Delta^{(2)} | bb J\rangle$ with $a=0p_{1/2}$, $b=0p_{3/2}$,  for $^{14}$C and $a=b=0d_{5/2}$ for $^{24}$Mg.
While we only show two specific matrix elements, a similar pattern is observed for the other matrix elements. 

Based on this observation, we may also understand the relative importance of the diagram topologies in Fig.~\ref{fig_topology}.
These may be ordered by the number of contractions between the two $\Omega$ operators.
$\Gamma^{\MyRomanNum{1}}$ has three contractions, $\Gamma^{\MyRomanNum{3}}$ has two contractions, $\Gamma^{\MyRomanNum{4}}$ has one contraction, and $\Gamma^{\MyRomanNum{2}}$ has zero contractions; this coincides with the ordering of the norms of the topologies in Fig.~\ref{fig_topology}.
Likewise, $f^{\MyRomanNum{1}}$ has three contractions, $f^{\MyRomanNum{3}}$ has two contractions, and $f^{\MyRomanNum{2}}$ has one contraction, again coinciding with their relative importance.

Finally, we note that the $f^{\MyRomanNum{3}}$ and $\Gamma^{\MyRomanNum{3}}$ topologies are significantly enhanced in the realistic case, as compared with the random operators, which indicates that this enhancement depends on the structure of the Hamiltonian.
It is suggestive that $f^{\MyRomanNum{3}}$ and $\Gamma^{\MyRomanNum{3}}$ have the form of the leading corrections to the self-energy and two-body vertex, respectively, due to a virtual one-phonon state in the random phase approximation (RPA).

To pursue this further, the bottom row of Fig.~\ref{fig_NormOfDoubleCommutators} shows results obtained with a schematic quadrupole-quadrupole interaction $V=Q\cdot Q$.
We find even greater enhancement of the $f^{\MyRomanNum{3}}$ and $\Gamma^{\MyRomanNum{3}}$ topologies than in the realistic case.
This enhancement with a separable interaction may be understood by considering the structure of the expression for $f^{\MyRomanNum{3}_a}$ given in Appendix~\ref{appendix:M-scheme}.
Since the $Q\cdot Q$ interaction is separable in the particle-hole channel, the particle-hole representation of the interaction has the form $\bar{\Gamma}_{i\bar{j}k\bar{l}}\sim Q_{ij}Q_{kl}$.
Assuming $\Omega$ also retains this form, we have
\begin{equation}
\begin{aligned}
    f_{ij}^{\MyRomanNum{3}_a} &\sim \sum_{abcde} \bar{\Omega}_{i\bar{c}d\bar{e}}\bar{\Omega}_{d\bar{e}a\bar{b}}\bar{\Gamma}_{a\bar{b}j\bar{c}}   \\
    &\sim \sum_{abcde} Q_{ic}Q_{de} Q_{de}Q_{ab} Q_{ab} Q_{jc}.
    \end{aligned}
\end{equation}
Especially for the diagonal terms $i=j$, the summand is a product of three squared quantities, which is manifestly positive, and we again find coherence in the sum.
Similar behavior exists for the $\Gamma^{\MyRomanNum{3}}$ topology, especially for the case $\Gamma_{ijij}$.
This coherence occurs only to the extent that the interaction is separable.
This suggests that a singular-value decomposition, in addition to providing an efficient truncation scheme for many-body operators~\cite{HergertSVD}, could be useful in identifying which channels and topologies might be enhanced.

To summarize: we find a hierarchy among the topologies in which diagrams with more contractions between the two $\Omega$ operators have a larger contribution, due to the coherent enhancement of the corresponding sums.
In addition, the specific structure of the nucleon-nucleon interaction, specifically the approximate separability corresponding to coherent states, can lead to additional enhancements.

\section{Application of IMSRG(\texorpdfstring{3${\rm f}_2$}{3f2}) to large-scale calculations}
We now turn to large-scale calculations of medium-mass nuclei.
In what follows, we use a single-particle space with $e_{\rm max}$ = 12 and additionally truncate the input 3N interaction to $E3_{\rm max}$ = 18.
The (dimensionless) threshold of the norm of $\Omega_H$ in the hunter-gatherer scheme is taken as 0.1.
Unless otherwise stated, the perturbative triples correction is included for the ground state energy.

\subsection{Comparison of single-reference and valence-space calculations}

\begin{figure}[ht]
    \centering
        \includegraphics[width=\linewidth]{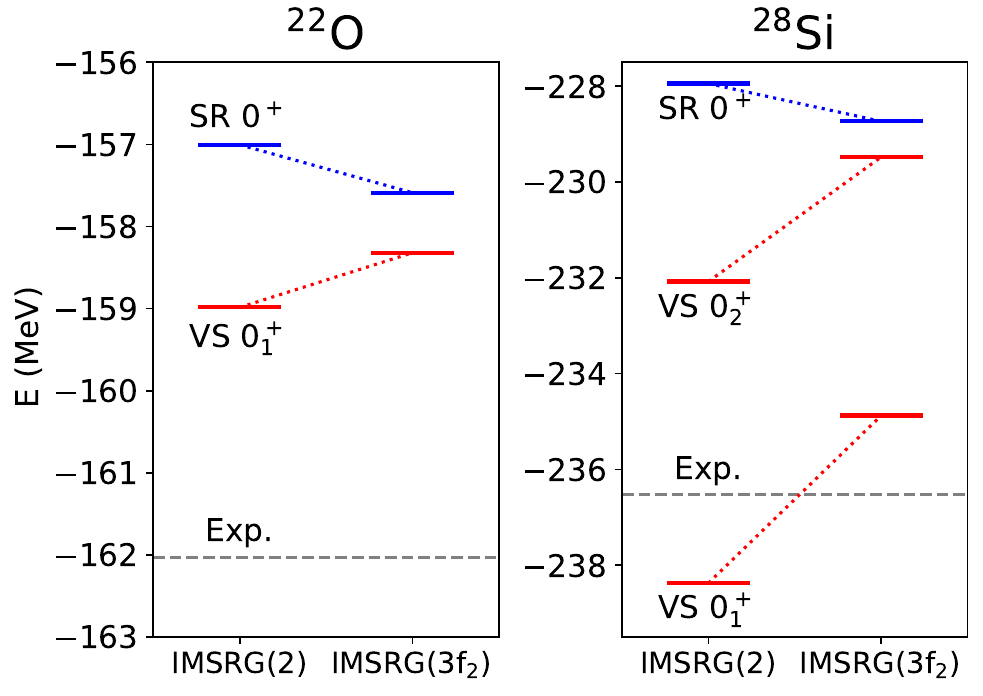}
    \caption{The calculated ground states of $^{22}$O and $^{28}$Si with valence space (VS) and single reference (SR) formulations of the IMSRG. For $^{28}$Si, we also include the excited $0^+_2$ state.
    }
    \label{fig_SingleRef}
\end{figure}

Closed-subshell nuclei provide a useful benchmark for the IMSRG because they can be computed both with the single-reference and valence space formulations.
In the limit where all induced operators are retained and the transformation is perfectly unitary, both formulations should give identical results.
When the induced operators are truncated and the transformation is only approximately unitary, the two methods will make different errors because they are performing different unitary transformations.
We therefore expect that an improved calculation should bring the results of the single-reference and valence space formulations into better agreement.

To test this expectation, we compute the ground states energies of $^{22}$O and $^{28}$Si with single-reference and valence-space formulations, using the IMSRG(2) and IMSRG($3f_2$) approximations.
The results are shown in Fig.~\ref{fig_SingleRef}.
For $^{22}$O, the discrepancy between the single-reference and valence-space energies is reduced from 2~MeV to 1~MeV when going from IMSRG(2) to IMSRG($3f_2$).
Likewise, the discrepancy for $^{28}$Si is reduced from 4~MeV to 1~MeV, if we interpret the single-reference calculation as targeting the first excited $0^+$ state rather than the ground state.
(This is plausible because in the valence space calculation, the closed-subshell $(d_{5/2})^{12}$ configuration makes up 9\% of the $0^+_1$ state wave function and 15\% of the $0^+_2$ wave function.)
This supports the idea that doing a more accurate calculation makes the transformation more unitary, and thus reduces dependence on how the Hamiltonian is partitioned.
In the future, this will be a valuable check when comparing the VS-IMSRG with the IM-GCM~\cite{Yao2018}, which has a significantly different partitioning scheme.

\subsection{Comparison with SMCC}

\begin{figure}[t]
    \centering
    \includegraphics[width=\columnwidth]{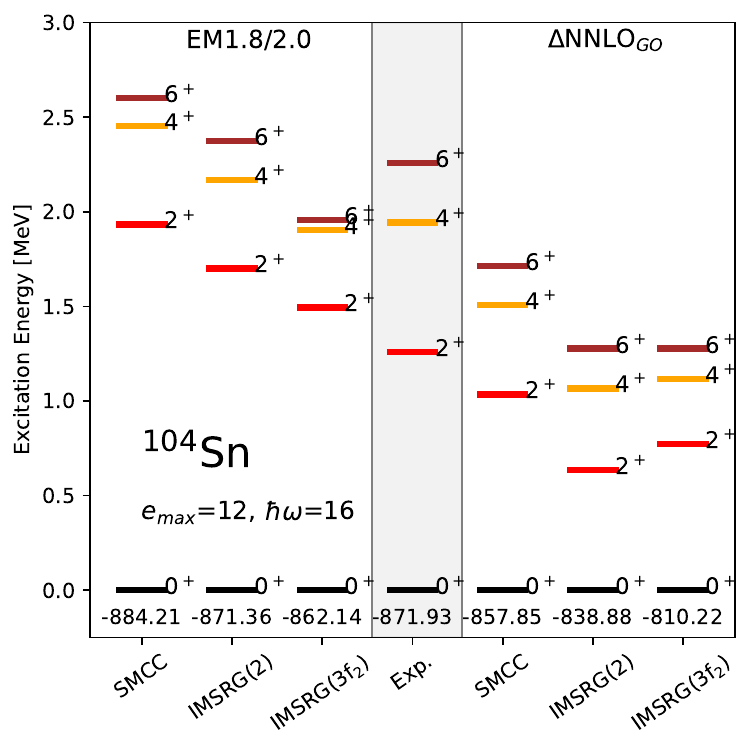}
    \caption{Excited state spectrum of $^{104}$Sn computed with different approximation schemes and two different chiral interactions. The SMCC results are from ref.~\cite{Sun2021}. The shaded band highlights the experimental spectrum~\cite{nndc_ensdf}. The absolute ground state energy, in MeV, is listed below the $0^+_1$ state for each spectrum.}
    \label{fig:Sn104}
\end{figure}

As mentioned in section~\ref{sec:Double_commutators}, the VS-IMSRG
has many similarities with the SMCC approach~\cite{Sun2018}.
Consequently, we might expect that the two methods should give similar results when using the same interaction.
SMCC calculations in the region around $^{100}$Sn were reported in ref.~\cite{Sun2021}, using two chiral interactions, EM1.8/2.0\cite{HEBELER20211} and $\Delta$NNLO$_{GO}$~\cite{Jiang2020}.
We compare the SMCC results with VS-IMSRG for $^{104}$Sn in Fig.~\ref{fig:Sn104}.
For both interactions, as in ref.~\cite{Sun2021}, we use $e_{\rm mas}=12$, and $E_{\rm 3max}=16$.
The specific truncation scheme used in the SMCC calculations is most closely analogous to IMSRG(2) with a perturbative treatment of $\Omega_{\rm 3b}$; we include the contribution of $\Omega_{\rm 3b}$ to the zero-body term in our perturbative triples correction.
For the  EM 1.8/2.0 interaction, the perturbative triples correction is approximately 6 MeV, so the ground state energy of $^{104}$Sn with IMSRG(2) plus perturbative triples is about -877 MeV, which differs from the SMCC result by about 7 MeV.
However, the nested commutators included in the IMSRG(3f$_2$) approximation shift the ground state energy, including perturbative triples, to -862 MeV, which differs from the SMCC result by about 22 MeV.

Interestingly, we find that the $2^+$ excitation energies obtained with SMCC and IMSRG(3f$_2$) differ by nearly 500 keV, which is larger than the shift from IMSRG(2) to IMSRG(3f$_2$) and comparable with the variation due to the input interaction.
Likewise, the ground state energy differs by over 20~MeV, which is again comparable with the interaction dependence.
It is important to note that in ref~\cite{Sun2021}, the 
the double commutator term $[[S_{\rm 2b},H_{\rm 2b}]_{\rm 3b}]_{\rm 2b}$ is not included in the particle-hole formulation.
This may be the origin of the observed discrepancy.

\subsection{Carbon, sulfur, and nickel isotopic chains}
In this section, we explore the impact of the IMSRG(3f$_2$) approximation on ground state energies and excitation spectra of carbon, sulfur and nickel isotopic chains. 
We adopt the $0\hbar\omega$ valence model space (i.e. a single major oscillator shell encompassing the naive Fermi level) for both protons and neutrons.
For example, for $^{38}$S we have valence protons in the $sd$ shell and valence neutrons in the $fp$ shell.
For a closed neutron shell, e.g. $^{36}$S, we do not decouple any valence neutron orbits.
We encountered some difficulty decoupling the $0\hbar\omega$ space for $^{15,16}$C, for which continuum effects drive the neutron $1s_{1/2}$ orbit down in energy, leading to an intruder-type problem~\cite{Stroberg2019} associated with including the $0d_{3/2}$ orbit in the valence space.
For IMSRG(2), removing the $0d_{3/2}$ orbit from the valence space enabled the $^{15}$C calculation to converge, while using a larger single-particle space with $e_{\rm max}$ enabled $^{16}$C to converge.
For IMSRG(3f$_2$), using a natural orbitals basis~\cite{Tichai2019} enabled convergence for both isotopes.
All calculations for the sulfur and nickel chains converge with the usual $0\hbar\omega$ valence space and Hartree-Fock reference in an $e_{\rm max}=12$ space.

At the outset, we should emphasize the potential pitfall of using a comparison with experimental data to evaluate the quality of our many-body method.
The exact solution of the Schr\"odinger equation using any interaction will not agree with the experimental data with arbitrary precision.
It is possible that a systematic error due to the interaction could roughly cancel with a systematic error in the many-body solution, giving the wrong impression about the accuracy of the method.
Nevertheless, given the remarkable performance of the EM 1.8/2.0 interaction for energies~\cite{Hagen2016,Simonis2017,Morris2018}, we will tacitly assume that discrepancies with the experimental data, particularly for excitation spectra, are due to errors in the many-body method.

\begin{figure}[ht]
    \centering
    \includegraphics[width=0.98\linewidth]{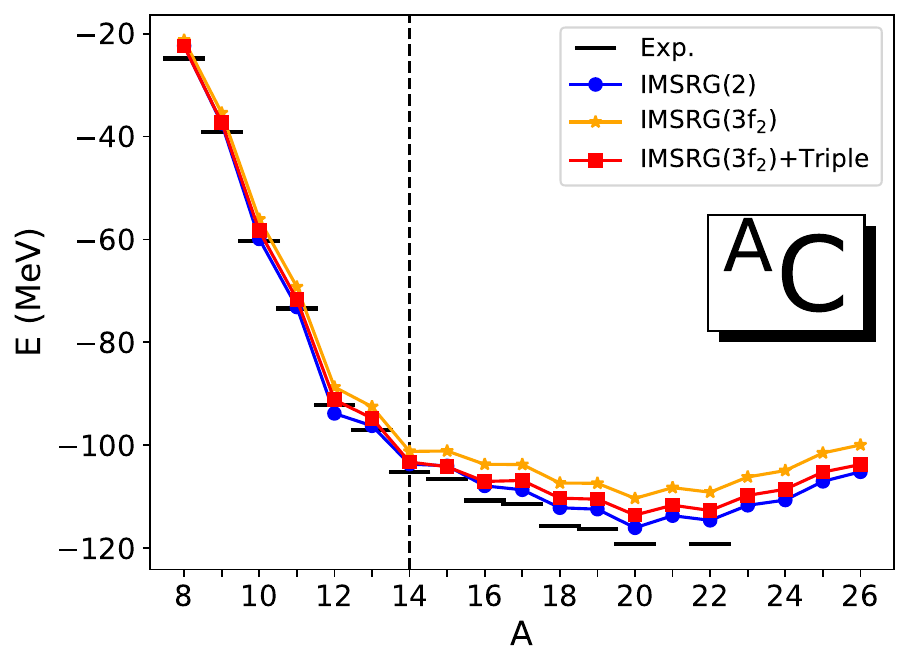}
    \includegraphics[width=0.98\linewidth]{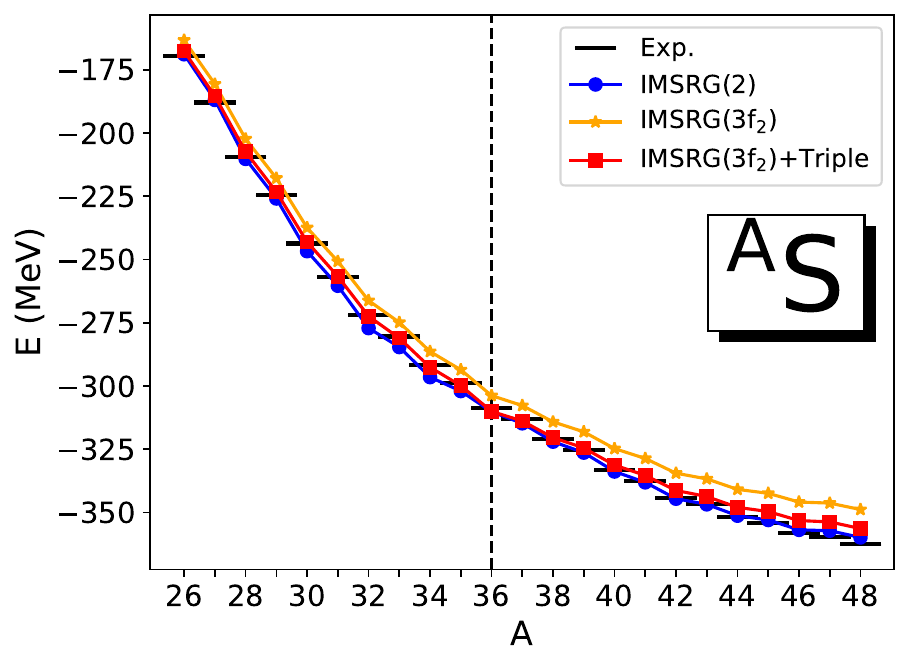}
    \includegraphics[width=0.98\linewidth]{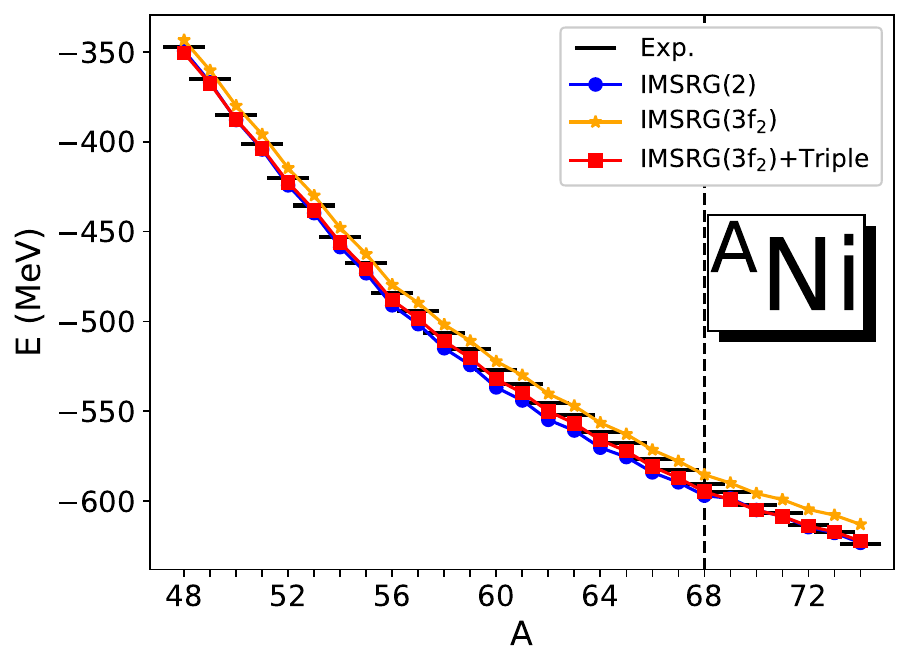}
    \caption{The calculated ground-state energies of Carbon, sulfur and Nickel isotopes chains compared with experiment (AME 2012, bars)~\cite{Wang_2021}. 
    }
    \label{fig_BindingE}
\end{figure}

The ground state energies of the C, S, and Ni isotopic chains are shown in the Fig. \ref{fig_BindingE}.
In general, the IMSRG(3f$_2$) energies are slightly higher compared to those from IMSRG(2).
For the carbon chain, the average shift due to IMSRG(3f$_2$) is +4.4 MeV,
while the average shift due to the perturbative triples is -2.9 MeV.
This approximate cancellation has been observed previously~\cite{Morris2016}.
Similar effects can be found in sulfur and nickel chain.
The upwards shift in the energy due to IMSRG(3f$_2$) can be understood based on the argument of section~\ref{sec:topologies}, in which the two-body matrix elements are generically reduced by the $\Gamma^{\MyRomanNum{1}}$ topology.
The diagonal matrix elements of the two-body interaction are generally attractive, and so reduction gives less attraction, while the suppression of the off-diagonal terms reduces the correlation energy.

Since IMSRG(2) already captures the binding energies very well, any discrepancies are hard to discern from the plot. 
We use the standard deviation of the shifts from IMSRG(2) to experimental results and from IMSRG(3f$_2$) plus triples to the experiment to illustrate the differences.
For the carbon chain, the standard deviation of binding energies for IMSRG(2) and IMSRG(3f$_2$) is 1.60 MeV and 1.78 MeV, respectively. 
For the sulfur chain, these values are 2.13 MeV for IMSRG(2) and 2.18 MeV for IMSRG(3f$_2$).
For the nickel chain, they are 3.09 MeV and 1.68 MeV, respectively.
Except for the nickel chain, IMSRG(3f$_2$) shows slightly worse performance in binding energies, but for the nickel chain, there is a clear improvement. 
However, this comparison is not sufficient to judge the overall effectiveness of IMSRG(3f$_2$). In ab initio calculations, directly comparing with experimental results is risky due to our incomplete understanding of the interactions.

\begin{figure}[ht]
    \centering
    \includegraphics[width=0.95\linewidth]{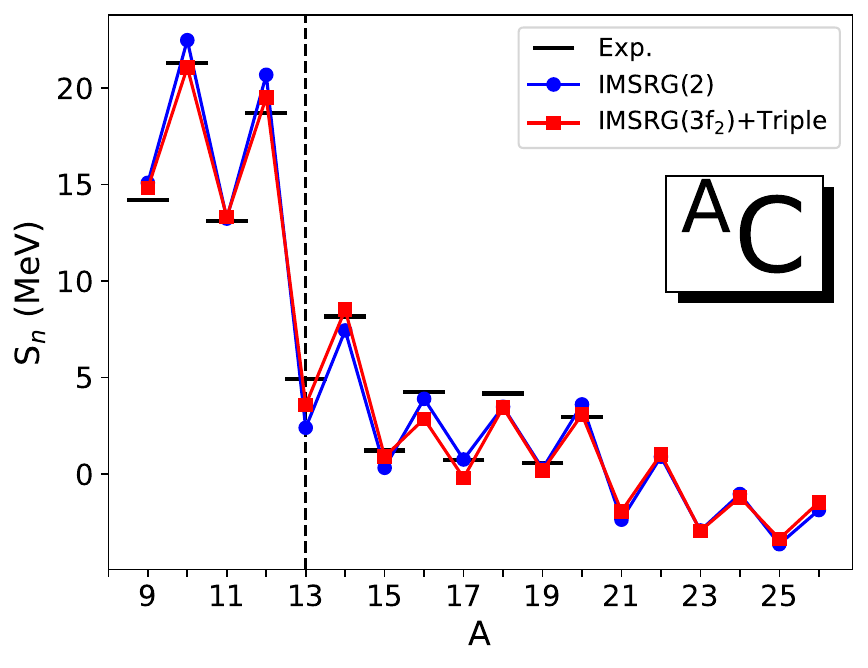}
    \includegraphics[width=0.95\linewidth]{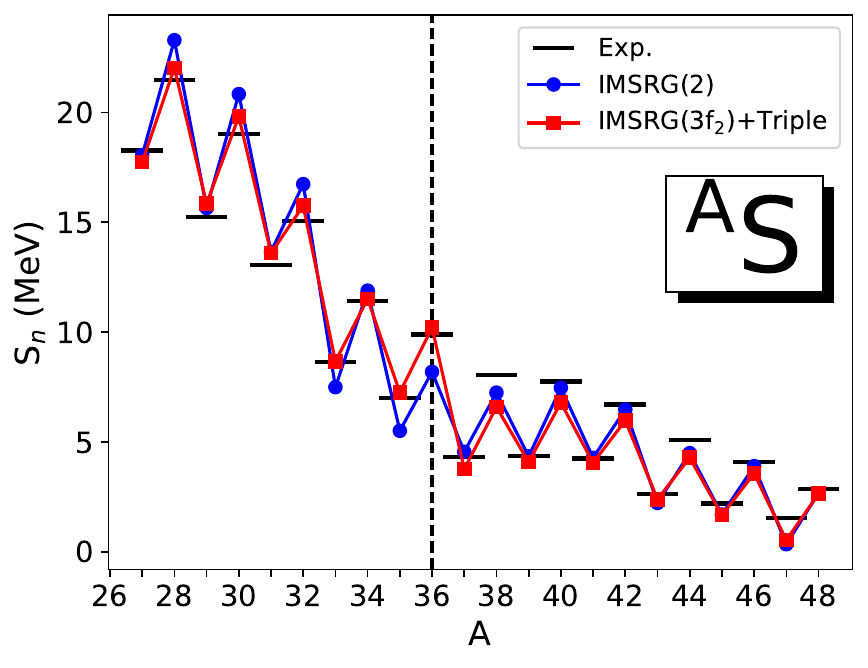}
    \includegraphics[width=0.95\linewidth]{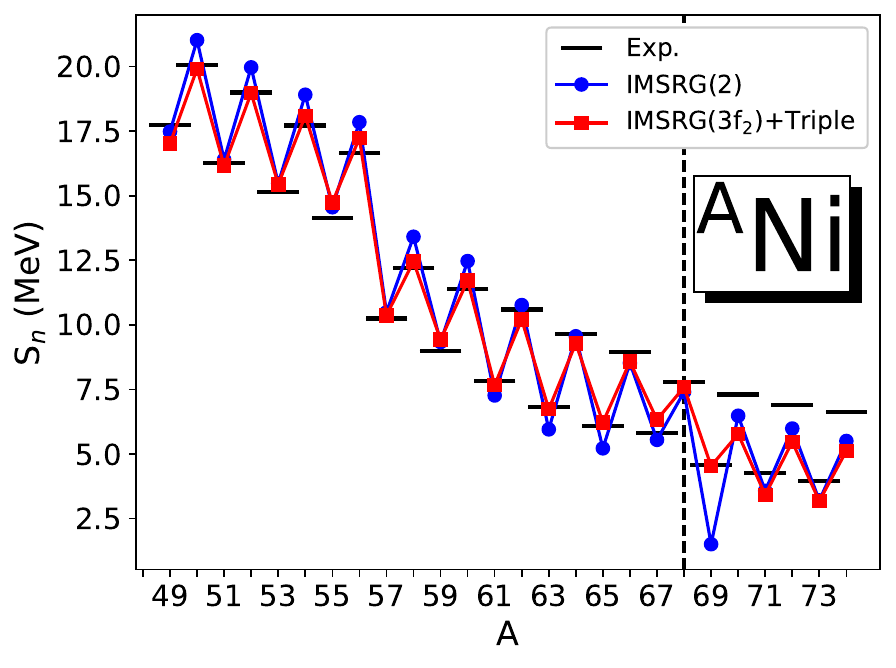}
    \caption{Computed single-neutron separation energies ($S_n$) of carbon, sulfur and nickel isotopes chains, compared with experiment. The dashed vertical lines indicate a change in valence space. 
    }
    \label{fig_Sn}
\end{figure}

\begin{figure}[ht]
    \centering
    \includegraphics[width=0.95\linewidth]{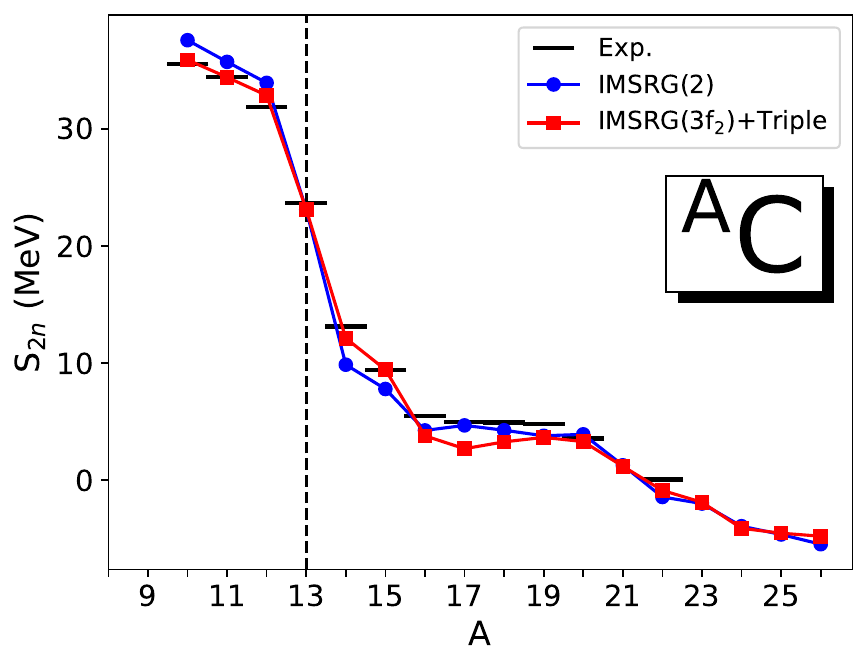}
    \includegraphics[width=0.95\linewidth]{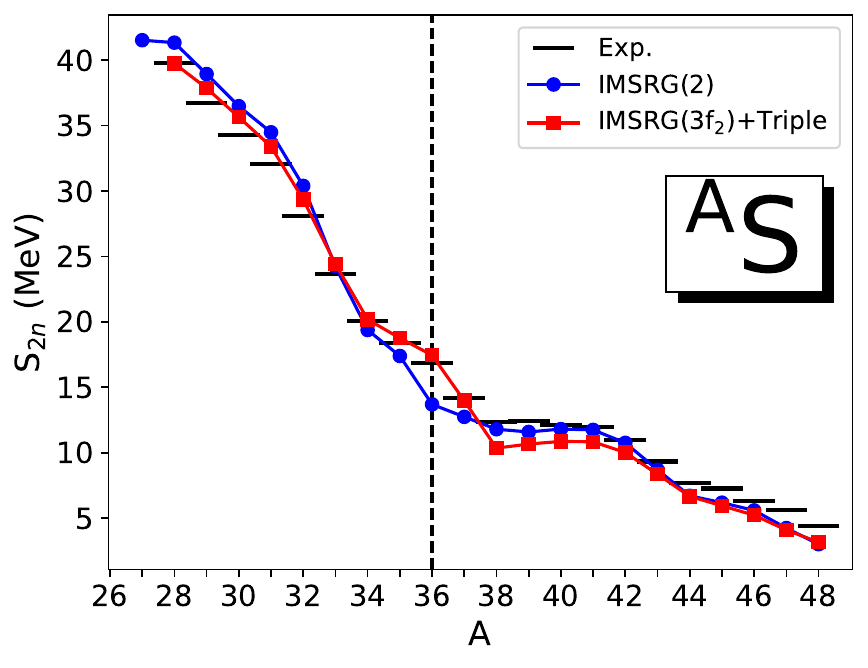}
    \includegraphics[width=0.95\linewidth]{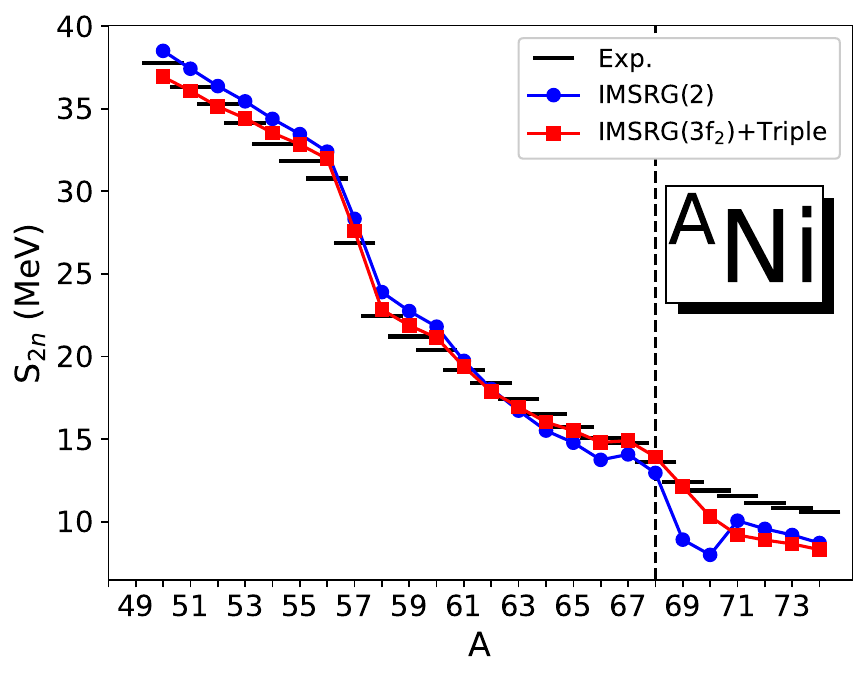}
    \caption{Computed two-neutron  separation energies ($S_{2n}$) of carbon, sulfur and nickel isotopes chains, compared with experiment. The dashed vertical lines indicate a change in valence space.
    }
    \label{fig_S2n}
\end{figure}

To further analyze the IMSRG(3f$_2$) results, the single-neutron and two-neutron separation energies are calculated and compared in Fig. \ref{fig_Sn} and \ref{fig_S2n}, respectively.
As we can see, the separation energies from IMSRG(2) diverge from the experimental values around the shell closure, while they reproduce the energies accurately at other locations. 
This discrepancy may be understood by considering the choice of valence space.
For a typical calculation of the separation energy, both nuclei entering the expression are computed in the same valence space.
Consequently, the transformation performed is essentially identical in the two cases and common systematic errors will cancel when taking the difference.
On the other hand, at the shell closure two different valence spaces are required to calculate the separation energy.
In this case, the IMSRG transformation is different for the two nuclei and the cancellation of errors will be less complete.
One-neutron separation energies $S_n$ of the C, S, and Ni chains are shown in Fig.~\ref{fig_Sn}, while two-neutron separation energies $S_{2n}$ are in Fig.~\ref{fig_S2n}.
The main effect of going from IMSRG(2) to IMSRG($3f_2$) is to remove the artifacts around the shell closures, indicated with the vertical dashed lines.
For the rest of the isotopic chains, the IMSRG(3f$_2$) results are close to those from IMSRG(2), and to the experimental data.
The root-mean-square deviations from experiment are given in Table~\ref{tab:rms}.
Based on this, it appears that IMSRG(3f$_2$) indeed improves the performance in predicting $S_n$ and $S_{2n}$, though we reiterate that one should be cautious when drawing conclusions by comparing with experimental results.

\begin{table}[t]
    \centering
    \caption{Root-mean-square deviation with respect to experiment of binding energies, S$_n$ and S$_{2n}$ for carbon, sulfur and nickel chain, in units of MeV. \label{tab:rms}}
    \begin{tabular}{l| ccc |ccc| ccc}
    \hline\hline
    &\multicolumn{3}{c|}{$E$} &
    \multicolumn{3}{c|}{$S_n$} &
    \multicolumn{3}{c}{$S_{2n}$} \\
    &C&S&Ni&
    C&S&Ni&
    C&S&Ni\\
    \hline
    IMSRG(2)      & 1.60 & 2.13 & 3.09 & 1.25 & 0.94 & 0.94 & 1.59 & 1.46 & 1.59   \\
    IMSRG($3f_2$) & 1.78 & 2.18 & 1.68 & 0.73 & 0.60 & 0.60 & 0.96 & 1.08 & 1.04 \\ 
    \hline\hline
    \end{tabular}
\end{table}

\begin{figure}[ht]
    \centering
        \includegraphics[width=0.90\linewidth]{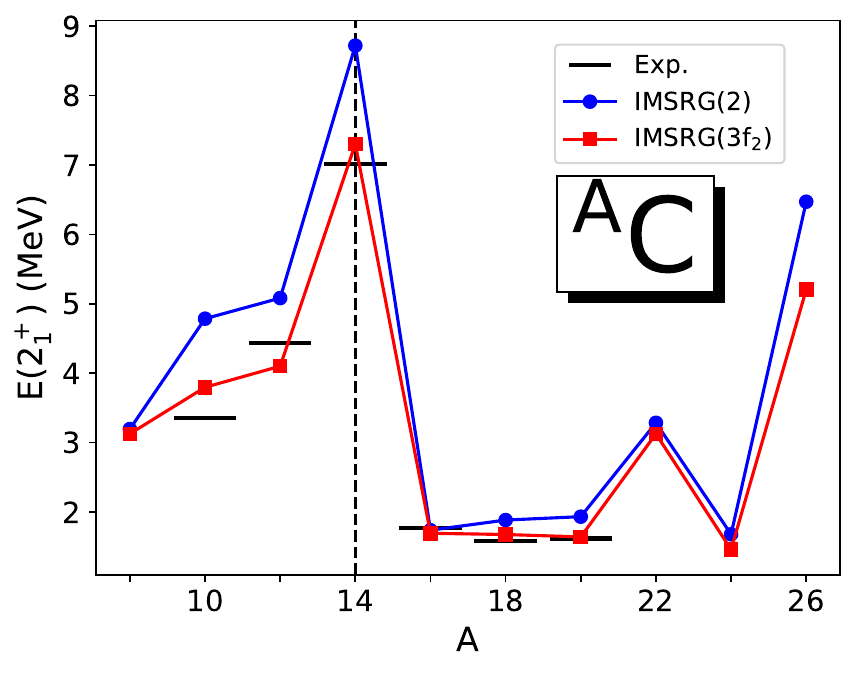}
        \includegraphics[width=0.90\linewidth]{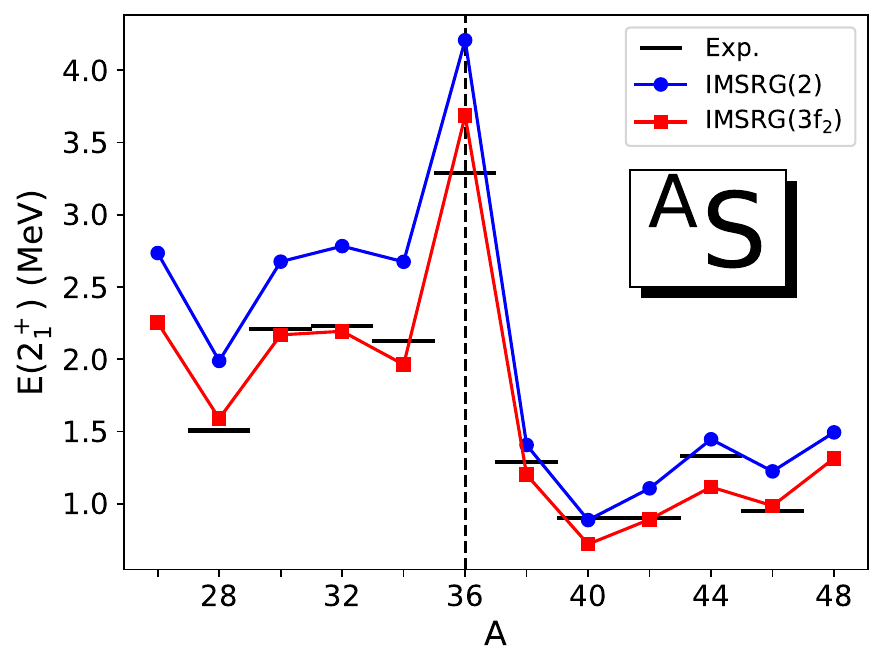}
        \includegraphics[width=0.90\linewidth]{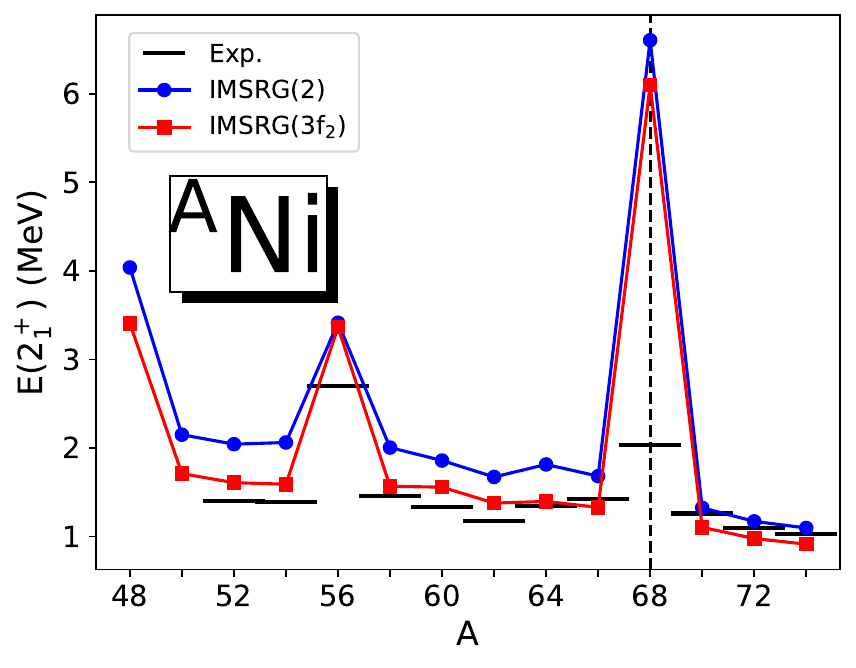}
    \caption{Computed excitation energy of the first 2$^+$ states of even carbon, sulfur and nickel isotopic chains, compared with experiment~\cite{nndc_ensdf}. }
    \label{fig_2+}
\end{figure}

For the nickel chain, the separation energies computed with both IMSRG(2) and IMSRG(3f$_2$) diverge from the experiments above $^{68}$Ni, reflecting an underbinding by about 1 MeV
per nucleon beyond $N=40$.
In a single-particle picture, this would reflect a gap between the $fp$ and $g_{9/2}$ orbits that is too large.
Such a gap would explain the small separation energies in Figs.~\ref{fig_Sn} and~\ref{fig_S2n}, as well as the very large $2^+$ excitation energy for $^{68}$Ni in Fig.~\ref{fig_2+}.
It is also consistent with previous findings in the $N=40$ region~\cite{Mougeot2018}.
As a further test, we decouple a valence space of the $\{f_{5/2},p_{3/2},p_{1/2},g_{9/2}\}$ neutron orbits on top of a $^{56}$Ni core, and compute the spectrum of $^{69}$Ni.
We find that the excited $\tfrac{1}{2}^-$ state in $^{69}$Ni is 2.55 MeV with IMSRG(2)(2.79~MeV with 3f$_2$) above the $\tfrac{9}{2}^+$ ground state, compared with the experimental value of 321 keV~\cite{nndc_ensdf}.
It remains unclear whether this gap is a feature of the interaction used, or if it reflects missing correlations (e.g. a static oblate deformation).
This point will be pursued in future work.

Perhaps the most significant improvement from the IMSRG(3f$_2$) comes in the $2^+_1$ excitation energies of even even nuclei, which are shown for the C, S, and Ni isotopes in Fig.~\ref{fig_2+}.
We obtain a significantly improved description across all three isotopic chains, with two notable exceptions in the nickel chain;
$^{56}$Ni, in which the $2^+_1$ energy is essentially unchanged, and $^{68}$Ni, where the $2^+_1$ energy is $\sim 6$~MeV compared with the experimental value of $\sim 2$~MeV.
As discussed above, the $^{68}$Ni excitation energy can be interpreted as reflecting a shell gap at $N=40$ that is too large.
(Note that in the $^{68}$Ni calculations shown in ~\ref{fig_2+}, only protons are active in the valence space.)

\section{summary}
In this paper, we have introduced an approximation to IMSRG(3), denoted IMSRG(3f$_2$) in which the effects of intermediate three-body operators are included at the level of two nested commutators.
This scheme maintains the $N^6$ scaling of the standard IMSRG(2) approximation, and can be applied to large-scale calculations.
We analyzed which terms in the IMSRG(3f$_2$) approximation are most important, and enhancement from coherent sums.
The coherence may either be driven by features of the interaction, or purely by the diagram topology.

Compared with IMSRG(2), the IMSRG(3f$_2$) generally produces more accurate spectroscopy, and largely eliminates artifacts in separation energies due to the change of valence space.
Two nickel isotopes, $^{56}$Ni and $^{68}$Ni, provide notable exceptions to the general improvement, and will be investigated in further work.

The approximation scheme presented here can in principle be extended to treat three nested commutators, though the number of diagrams required likely makes a direct implementation prohibitive.
Building on the understanding of enhancement mechanisms discussed in this work may help to select the most relevant contributions.
Another straightforward extension will be to generalize the $J$ coupled expressions presented here to spherical tensor operators.
This will be especially interesting to investigate the impacts on electric quadrupole observables which are generally underpredicted at the IMSRG(2) level~\cite{Parzuchowski2017a,Stroberg2022}.

\begin{acknowledgments}
We would like to thank Titus Morris and Zonghao Sun for helpful discussions regarding the SMCC.
This research was supported in part by the Notre Dame’s Center for Research Computing.
\end{acknowledgments}

\appendix

\section{M-scheme double commutators}
\label{appendix:M-scheme}
The one-body piece of the double commutator is
\begin{subequations}\allowdisplaybreaks[4]  \label{DoubleCommutator_onebody_all}
\begin{align}
%%%%%%%%%%%%%%%%%  I
 f^\MyRomanNum{1}_{ij}  =&\frac{1}{2} \sum_{abcde} \left( \bar{n}_a \bar{n}_b  {n}_c  {n}_d  -  {n}_a  {n}_b \bar{n}_c  \bar{n}_d  \right)   \nonumber \\
  &\times \left( \Omega_{cdab}\Omega_{abce}   \Gamma_{eidj}  +  \Omega_{cdab} \Omega_{abce}  \Gamma_{diej}  \right)\\
 %%%%%%%%%%%%%%%%  II   
  f^{\MyRomanNum{2}}_{ij}  =& \frac{1}{2} \sum_{abcde} \left( {n}_a  {n}_b  \bar{n}_c  \bar{n}_e  -   \bar{n}_a  \bar{n}_b  {n}_c  {n}_e  \right) \nonumber \\
 &\times  \left( \Gamma_{cdab} \Omega_{abce}   \Omega_{eidj}   - \Gamma_{cdab} \Omega_{abce}  \Omega_{diej}  \right)  \\
f^{\MyRomanNum{3}_a}_{ij} =& \sum_{abcde} \left( \bar{n}_a  \bar{n}_b  {n}_c  {n}_d  -  {n}_a  {n}_b  \bar{n}_c  \bar{n}_d \right) \nonumber \\
 &\times   \left( \Omega_{abcd}  \Omega_{idae} \Gamma_{cejb}  -  \Omega_{abcd}  \Omega_{edaj} \Gamma_{cieb}  \right)\\
f^{\MyRomanNum{3}_b}_{ij}  =& \frac{1}{4} \sum_{abcde} \left( \bar{n}_a  \bar{n}_b  {n}_c {n}_d  -  {n}_a  {n}_b  \bar{n}_c  \bar{n}_d \right) \nonumber \\
 &\times   \left( \Omega_{abcd} \Omega_{cdej} \Gamma_{eiab} -  \Omega_{abcd} \Omega_{eiab} \Gamma_{cdej} \right)
\end{align}
\end{subequations}

These may be re-expressed in a factorized form
\begin{subequations}\label{doubleCommutator_1b}
\begin{align} \allowdisplaybreaks[4] 
 f^\MyRomanNum{1}_{ij}  =& \sum_{ab} \left( \chi^{\alpha}_{ab}  \Gamma_{biaj} + \chi^{\alpha}_{ab} \Gamma_{aibj} \right) \\
   %%%%%%%%%%
f^{\MyRomanNum{2}}_{ij}   =& \sum_{ab} \left( \chi^{\beta}_{ab} \Omega_{biaj} - \chi^{\beta}_{ab} \Omega_{aibj} \right)\\
  %%%%%%%%%%
 f^{\MyRomanNum{3}_a}_{ij}  =& \sum_{abc} \left( \chi^{\gamma}_{icab} \Gamma_{abjc} -  \Gamma_{ciab}  \chi^{\gamma}_{abcj} \right)\\
  %%%%%%%%%%
f^{\MyRomanNum{3}_b}_{ij}  =& \sum_{abc} \left(\chi^{\delta}_{ciab}  \Gamma_{abcj} - \Gamma_{ciab} \chi^{\delta}_{abcj} \right)
\end{align}
\end{subequations}
where we have defined the intermediate quantities
\begin{subequations}\label{chi_1b}
\begin{align} \allowdisplaybreaks[4]
\chi^{\alpha}_{ij} =&\frac{1}{2} \sum_{abc} \left( \bar{n}_a \bar{n}_b  {n}_c  {n}_i  -  {n}_a {n}_b \bar{n}_c  \bar{n}_i \right)  \Omega_{ciab} \Omega_{abcj}  \\
%%%%%%%%%%%%%%%%%%%%%%%%%%%%%%%%%%%%%%%%%%%%%%%%%%%%%%%%%%%%%%%%
\chi^{\beta}_{ij} =&\frac{1}{2} \sum_{abc} \left(  \bar{n}_a   \bar{n}_b {n}_c {n}_i  -  {n}_a  {n}_b   \bar{n}_c   \bar{n}_i \right) \Omega_{ciab} \Gamma_{abcj} \\
%%%%%%%%%%%%%%%%%%%%%%%%%%%%%%%%%%%%%%%%%%%%%%%%%%%%%%%%%%%%%%%%
\chi^{\gamma}_{ijkl} =& \sum_{ab} \left(  {n}_a \bar{n}_b {n}_j  \bar{n}_k   -  \bar{n}_a {n}_b \bar{n}_j  {n}_k   \right)  \Omega_{ajkb} \Omega_{ibal} \\
\chi^{\delta}_{ijkl} =& \frac{1}{4}\sum_{ab} \left( \bar{n}_i \bar{n}_j {n}_a {n}_b  -  {n}_i  {n}_j  \bar{n}_a \bar{n}_b  \right)  \Omega_{ijab} \Omega_{abkl}
\end{align}
\end{subequations}
We use superscript Greek letters to distinguish the intermediates.

The two-body piece of double the commutator is
\begin{subequations}\allowdisplaybreaks[4]\label{DoubleCommutator_twobody_all}
\begin{align}
%  Ia
\Gamma^{\MyRomanNum{1}}_{ijkl}  =&\frac{1}{2} \sum_{abcd} \left( \bar{n}_a  \bar{n}_b  {n}_c  +  {n}_a  {n}_b  \bar{n}_c \right) \nonumber\\
& \times \left\{ (1\! -\! \hat P_{ij} ) \Omega_{ciab} \Omega_{abcd} \Gamma_{djkl} \right. \nonumber\\
&~+ \left.  (1\! -\! \hat P_{kl} ) \Omega_{cdab} \Omega_{abcl} \Gamma_{ijkd} \right\}\\
%    IV
\Gamma^{\MyRomanNum{2}}_{ijkl}   =& 
- \frac{1}{2} \sum_{abcd} \left(   \bar{n}_a \bar{n}_b {n}_c +  \bar{n}_c  {n}_a  {n}_b \right)  \nonumber \\
&\times  \left\{  (1\! -\! \hat P_{ij} )  \Omega_{cjab}  \Gamma_{abcd}  \Omega_{idkl}\right. \nonumber\\
&~+ \left.  (1\! -\! \hat P_{kl} ) \Gamma_{cdab}  \Omega_{abcl} \Omega_{ijkd}  \right\}\\
%   IIa
\Gamma^{\MyRomanNum{3}_a}_{ijkl} =& - \sum_{abcd} \left( \bar{n}_c  \bar{n}_d  {n}_a  +  {n}_a  \bar{n}_c  \bar{n}_d \right)  \nonumber\\
& \times \left\{  (1\! -\! \hat P_{ij} )  \Omega_{ajcd} \Omega_{idab} \Gamma_{cbkl} \right. \nonumber\\
& ~+ \left. (1\! -\! \hat P_{kl} ) \Omega_{cdka} \Omega_{bacl} \Gamma_{ijbd} \right\} \\
%   IIb
\Gamma^{\MyRomanNum{3}_b}_{ijkl}  =& - \sum_{abcd} ( \bar{n}_b  {n}_c  {n}_d  +  {n}_b  \bar{n}_c  \bar{n}_d ) (1 - \hat P_{ij} ) (1 - \hat P_{kl} )  \nonumber\\
& \times \left( \Omega_{dcbk} \Omega_{biac} \Gamma_{jald} + 
 \Omega_{jcbd} \Omega_{balc} \Gamma_{diak} \right) \\
%    IIc
\Gamma^{\MyRomanNum{3}_c}_{ijkl}  =& - \frac{1}{2} \sum_{abcd} (   \bar{n}_a \bar{n}_b {n}_c + {n}_a  {n}_b  \bar{n}_c)  
 (1 - \hat P_{ij} )  (1 - \hat P_{kl} )  \nonumber\\
 & \times  \left(  \Omega_{abcl} \Omega_{idab} \Gamma_{cjkd} +  \Omega_{icab} \Omega_{abdl} \Gamma_{djkc}   \right)\\
%    IIIb
\Gamma^{\MyRomanNum{4}_a}_{ijkl}  =&  -   \sum_{abcd} \left(   \bar{n}_c \bar{n}_d {n}_a  +  {n}_c  {n}_d  \bar{n}_a \right) \nonumber \\
& \times \left\{  (1\! -\! \hat P_{ij} ) \Omega_{aicd} \Omega_{dbkl} \Gamma_{jcba} \right. \nonumber \\
&+ \left.  (1 \!-\! \hat P_{kl} )  \Omega_{dcak} \Omega_{ijcb} \Gamma_{bald} \right\}\\
%    IIIa
\Gamma^{\MyRomanNum{4}_b}_{ijkl}  =&  (1 - \hat P_{ij} ) (1 - \hat P_{kl} ) \sum_{abcd} (   \bar{n}_a {n}_b \bar{n}_c +  {n}_a  \bar{n}_b  {n}_c )   \nonumber\\
& \times \left(\Omega_{bica} \Omega_{jcld} \Gamma_{dabk}  + \Omega_{cabk} \Omega_{jdlc} \Gamma_{bida} \right)\\
%    IIIc
\Gamma^{\MyRomanNum{4}_c}_{ijkl} =&   \frac{1}{2} (1 - \hat P_{ij} ) (1 - \hat P_{kl}) \sum_{abcd} (   \bar{n}_a \bar{n}_b {n}_d +{n}_a  {n}_b  \bar{n}_d ) \nonumber \\
&\times \left( \Omega_{abld} \Omega_{djck} \Gamma_{icab}  +\Omega_{idab} \Omega_{cjdk} \Gamma_{ablc}   \right)
\end{align}
\end{subequations}

These may be expressed in a factorized form
\begin{subequations}\allowdisplaybreaks[4]\label{Factorized_DoubleCommutator_twobody}
\begin{align}
\Gamma^{\MyRomanNum{1}}_{ijkl}  =& \sum_{a} \left\{(1\! -\! \hat P_{ij} )   \chi^{\epsilon}_{ia} \Gamma_{ajkl} % \right. \nonumber\\
+ (1\! -\! \hat P_{kl} )  \chi^{\epsilon}_{ak} \Gamma_{ijal}    \right\}\\
 %&+ \left. \left(1 - \hat P_{kl} \right)  \chi^{\epsilon}_{ak} \Gamma_{ijal}    \right\}\\
%
\Gamma^{\MyRomanNum{2}}_{ijkl}  =&\sum_{a}  \left\{
(1 \!-\! \hat P_{ij} )   \chi^{\zeta}_{aj}  \Omega_{iakl} % \right.\nonumber\\
%& - \left.
-(1 \!-\! \hat P_{kl} )  \chi^{\zeta}_{ak}  \Omega_{ijal}  \right\}\\
\Gamma^{\MyRomanNum{3}_a}_{ijkl}  =& 
-\!\sum_{ab} \bigl\{ (1\! -\! \hat P_{ij} ) \chi^{\eta}_{ijab}  \Gamma_{abkl} %\nonumber\\
  + (1 \!- \!\hat P_{kl} )  \Gamma_{ijab}  \chi^{\eta}_{abkl} \bigr\} \\
%-\left(1 - \hat P_{ij} \right) \sum_{ab}\chi^{\eta}_{ijab}  \Gamma_{abkl} \nonumber\\
%& - \left(1 - \hat P_{kl} \right) \sum_{ab}  \Gamma_{ijab}  \chi^{\eta}_{abkl}  \\
%
\Gamma^{\MyRomanNum{3}_b}_{ijkl} =& - (1\! -\! \hat P_{ij}) (1\! -\! \hat P_{kl} ) % \nonumber\\
%&\times
\sum_{ab} \left( \chi^{\eta}_{bkai} \Gamma_{jbla} + \chi^{\eta}_{lajb}  \Gamma_{aibk}  \right)\\
\Gamma^{\MyRomanNum{3}_c}_{ijkl}  =& - \frac{1}{2} \left(1 - \hat P_{ij} \right)  (1\! - \!\hat P_{kl} ) \sum_{ab}  \chi^{\theta}_{iabl} \Gamma_{bjka}   \\
\Gamma^{\MyRomanNum{4}_a}_{ijkl}  =&  -\!\sum_{ab} \left\{  (1\!- \!\hat P_{ij} ) \chi^{\kappa}_{ijab} \Omega_{bakl}  \right. %\nonumber\\ 
%&+ 
-\left.  (1\! -\! \hat P_{kl} ) \Omega_{ijab} \chi^{\kappa}_{klba} \right\}  \\
\Gamma^{\MyRomanNum{4}_b}_{ijkl}  =&  (1 \!- \!\hat P_{ij} )  (1 \!-\! \hat P_{kl} ) %\nonumber \\
%&\times 
\sum_{ab} \left( \chi^{\iota}_{aibk} \Omega_{jbla} - \chi^{\iota}_{akbi} \Omega_{jalb}  \right ) \\
%
%\Gamma^{\MyRomanNum{4}_c}_{ijkl}   =&  - \frac{1}{2} \left(1 - \hat P_{ij} \right) \left(1 - \hat P_{kl} \right)  \sum_{ab} \chi^{\lambda}_{jalb}  \Omega_{biak} 
\Gamma^{\MyRomanNum{4}_c}_{ijkl}   =&   \frac{1}{2} (1\! -\! \hat P_{ij} ) 
 (1 \!-\! \hat P_{kl} )  \sum_{ab} \chi^{\lambda}_{ialb}  \Omega_{bjak} 
\end{align}
\end{subequations}
where we have defined additional intermediates
\begin{subequations}\allowdisplaybreaks[4]\label{chi_2b}
\begin{align}
\chi^{\epsilon}_{ij} =& \frac{1}{2} \sum_{abc} \left( \bar{n}_a \bar{n}_b {n}_c    +  {n}_a  {n}_b \bar{n}_c \right)  \Omega_{ciab} \Omega_{abcj} \\  % I   ab
\chi^{\zeta}_{ij}  =&\frac{1}{2} \sum_{abc} \left(    {n}_a {n}_b \bar{n}_c + \bar{n}_a  \bar{n}_b {n}_c \right) \Gamma_{aibc}  \Omega_{bcaj} \\   % IV ab
\chi^{\eta}_{ijkl} =& \sum_{ab} \left( \bar{n}_a  {n}_b  \bar{n}_k  +  {n}_a  \bar{n}_b  {n}_k \right)  \Omega_{iabl} \Omega_{bjka} \\    % II  ac   % II  bd
\chi^{\theta}_{ijkl}  =&  \sum_{ab} \left(   {n}_a {n}_b \bar{n}_k  +  \bar{n}_a  \bar{n}_b {n}_k \right. \nonumber\\
& ~~~+ \left.   {n}_a {n}_b \bar{n}_j + \bar{n}_a  \bar{n}_b {n}_j  \right)   \Omega_{ijab} \Omega_{abkl}  \\   % II  ef
%%%%%%%%%%%%%%%%%%%%%%%%%%%%%%%%%%%%%%%%%%%%%%%%%%%%%%%%%%%%%%%
\chi^{\iota}_{ijkl} =& \sum_{ab} \left(   \bar{n}_a {n}_b \bar{n}_k  +  {n}_a  \bar{n}_b  {n}_k \right)  \Omega_{bika} \Gamma_{iabl} \\     % III ab
\chi^{\kappa}_{ijkl}  =&\sum_{ab} \left(   \bar{n}_a {n}_b {n}_k  +  {n}_a  \bar{n}_b  \bar{n}_k \right)  %\nonumber \\
\Omega_{ajbl} \Gamma_{ibka} \\     % III cd
\chi^{\lambda}_{ijkl}  =& \sum_{ab} \bigl\{ \left(   \bar{n}_a \bar{n}_b {n}_l +  {n}_a  {n}_b  \bar{n}_l \right)  \Gamma_{ijab} \Omega_{abkl}  \nonumber\\
                        &~~+   \left(   \bar{n}_a \bar{n}_b {n}_j +  {n}_a  {n}_b  \bar{n}_j \right)  \Omega_{ijab} \Gamma_{abkl} \bigr\}
% III ef
\end{align}
\end{subequations}
We note that the intermediates $\chi$ do not in general have Hermitian symmetry or permutation symmetry under exchange of indices.

\section{J-coupled factorization version of double commutators}
\label{section::Jscheme_factorizedDCs}
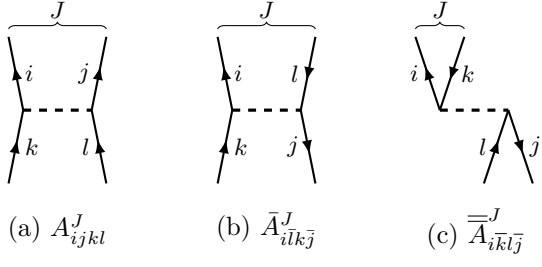
\begin{figure}[ht]
\centering
\begin{tikzpicture}
\node (standard){
 \begin{tikzpicture}[x=2em,y=3em]

\coordinate (OUT1) at (-1,+1);
\coordinate (OUT2) at (+1,+1);
\coordinate (IN1) at  (-1,-1);
\coordinate (IN2) at  (+1,-1);
\coordinate (V1) at (-0.7,0);
\coordinate (V2) at (+0.7,0);
%!!!!!!!!! Operator

\draw [fermion] (IN1) to node[right]{$k$} (V1);
\draw [fermion] (V1) to node[right]{$i$} (OUT1);
\draw [fermion] (IN2) to node[left]{$l$} (V2);
\draw [fermion] (V2) to node[left]{$j$} (OUT2);
\draw [boson] (V1) to (V2);

\draw[decoration={brace,mirror,raise=3pt},decorate] (OUT2) to node[above,yshift=4pt]{$J$} (OUT1);

\end{tikzpicture}
};

\node (pandya) at (standard.east) [xshift=20mm] {
 \begin{tikzpicture}[x=2em,y=3em]

\coordinate (OUT1) at (-1,+1);
\coordinate (OUT2) at (+1,+1);
\coordinate (IN1) at  (-1,-1);
\coordinate (IN2) at  (+1,-1);
\coordinate (V1) at (-0.7,0);
\coordinate (V2) at (+0.7,0);
%!!!!!!!!! Operator

\draw [fermion] (IN1) to node[right]{$k$} (V1);
\draw [fermion] (V1) to node[right]{$i$}  (OUT1);
\draw [fermion] (OUT2) to node[left]{$l$}(V2);
\draw [fermion] (V2) to node[left]{$j$} (IN2);
\draw [boson] (V1) to (V2);

\draw[decoration={brace,mirror,raise=3pt},decorate] (OUT2) to node[above,yshift=4pt]{$J$} (OUT1);

\end{tikzpicture}
};

\node (cross) at (pandya.east) [xshift=20mm] {
 \begin{tikzpicture}[x=2em,y=3em]

\coordinate (OUT1) at (-1.2,+1);
\coordinate (OUT2) at (-0.2,+1);
\coordinate (IN1) at  (0.2,-1);
\coordinate (IN2) at  (+1.2,-1);
\coordinate (V1) at (-0.7,0);
\coordinate (V2) at (+0.7,0);
%!!!!!!!!! Operator

\draw [fermion] (OUT2) to node[right]{$k$} (V1);
\draw [fermion] (V1) to node[left]{$i$} (OUT1);
\draw [fermion] (V2) to node[right]{$j$} (IN2);
\draw [fermion] (IN1) to node[left]{$l$} (V2);
\draw [boson] (V1) to (V2);

\draw[decoration={brace,mirror,raise=3pt},decorate] (OUT2) to node[above,yshift=4pt]{$J$} (OUT1);

\end{tikzpicture}
};
\node (a) at (standard.south) [yshift=-5mm,font=\normalsize] {(a) $A_{ijkl}^{J}$};
\node (b) at (pandya.south) [yshift=-5mm,font=\normalsize] {(b) $\bar{A}_{i\bar{l}k\bar{j}}^{J}$};
\node (c) at (cross.south) [yshift=-5mm,font=\normalsize] {(c) $\overline{\overline{A}}_{i\overline{k}l\overline{j}}^{J}$};

\end{tikzpicture}
\caption{\label{fig_coupling}Coupling schemes used in $J$-coupled expressions. (a)~Standard coupling, (b)~Pandya-transformed matrix elements, (c)~cross-coupled matrix elements.}
\end{figure}

To take advantage of rotational symmetry, we work with J-coupled matrix elements.
We work with un-normalized coupled states so that
\begin{equation}
    A_{ijkl}^{JM} = \sum_{m_im_jm_im_l} \mathcal{C}^{JM}_{j_im_i,j_jm_j}
     \mathcal{C}^{JM}_{j_km_k,j_lm_l}
    A_{ijkl}
\end{equation}
where the $\mathcal{C}$ are Clebsch-Gordan coefficients, and the uncoupled matrix element $A_{ijkl}$ depends on the projections $m_i,m_j,m_k,m_l$.
For the rotationally invariant operators considered in this work, $A^{JM}_{ijkl}$ is independent of the total projection $M$ and so $M$ is not explicitly indicated.
These expressions were obtained with the aid of the \texttt{amc} code~\cite{Tichai2020}.
In several cases, the expressions are simplified and made amenable to matrix multiplication if we use different coupling schemes.
In addition to the standard coupling scheme, we employ both Pandya-transformed and cross-coupled matrix elements, illustrated in Fig.~\ref{fig_coupling}.
Pandya-transformed matrix elements are indicated with a single bar
\begin{align} \label{Pandya}
\bar A^J_{i \bar l k \bar j} = - \sum_{J^\prime} (\hat{J}^\prime)^2 \sixj{{j}_{i}}{{j}_{l}}{{J}}{{j}_{k}}{{j}_{j}}{{J}^\prime}  A^{J^\prime}_{i j k l} 
\end{align}
where the braces indicate the six-J coefficient,
$j_a$ denotes the angular momentum of the orbit $a$
and we use the usual notation $\hat J \equiv \sqrt{2J+1}$.
Cross-coupled matrix elements are indicated with a double bar
\begin{align} \label{cross-coupled_Pandya}
 \overline{\overline{A}}^J_{j \bar l k \bar i} &= \sum_{J^\prime} (\hat{J}^\prime)^2 \sixj{{j}_{j}}{{j}_{l}}{{J}}{{j}_{k}}{{j}_{i}}{{J}^\prime}  (-1)^{j_i +j_j -J^\prime}   A^{J^\prime}_{i j k l} \nonumber \\
 \overline{\overline{A}}^J_{i \bar k l \bar j} &= \sum_{J^\prime} (\hat{J}^\prime)^2 \sixj{{j}_{i}}{{j}_{k}}{{J}}{{j}_{l}}{{j}_{j}}{{J}^\prime}  (-1)^{j_k +j_l -J^\prime}   A^{J^\prime}_{i j k l} 
\end{align}
In these definitions, we have not assumed Hermitian or permutation symmetries in the matrix elements $A$.

The one-body matrix elements are
\begin{subequations}\label{doubleCommutator_1b_Jcoupled_factorized}
\begin{align} \allowdisplaybreaks[4] 
 f^\MyRomanNum{1}_{ij}  =& \delta_{ j_i j_j} \hat{{j}}_{i}^{-2}  \sum_{ab J}  \hat{{J}}^{2}  \*  \chi^{\alpha}_{ab}  \Gamma^{J}_{biaj}      \\
 f^{\MyRomanNum{2}}_{ij}  =& \delta_{ j_i j_j} ~\hat j_i^{-2}  \sum_{ab J}  \hat J^2 \left(\chi^{\beta}_{ab}  -   \chi^{\beta}_{ba}  \right) \Omega^{J}_{biaj}   \\
 f^{\MyRomanNum{3}_a}_{ij}  =&  \delta_{ j_i j_j}~\hat j_i^{-2} \sum_{abc J} \left(  \bar \chi^{\gamma~J}_{i \bar c a \bar b} \Gamma^{J}_{a \bar b j \bar c} - \bar \chi^{\gamma~J}_{c \bar j a \bar b} \Gamma^{J}_{a \bar b c \bar i} \right)  \\
 f^{\MyRomanNum{3}_b}_{ij}  =& \delta_{ j_i j_j} ~\hat j_i^{-2} \sum_{abc J} \left(\chi^{\delta~J}_{ciab}  \Gamma^{J}_{abcj} - \chi^{\delta~J}_{abcj} \Gamma^{J}_{ciab} \right)
\end{align}
\end{subequations}
with intermediates
\begin{subequations}\label{Jcoupled_chi_1b}
\begin{align} \allowdisplaybreaks[4] 
 &\chi^{\alpha}_{ij} = \!  \sum_{abc J} \frac{\hat{J}^2}{2\hat{j}_{i}^{2}}   \left( \bar{n}_a  \bar{n}_b  {n}_c {n}_i  -  {n}_a  {n}_b  \bar{n}_c  \bar{n}_i \right) %\nonumber\\
  \Omega^{J}_{ciab} \Omega^{J}_{abcj} \\
%%%%%%%%%%%%%%%%%%%%%%%%%%%%%%%%%%%%%%%%%%%%%%%%
%&\chi^\beta_{ij} =\! \sum_{abc J} \frac{\hat{J}^2}{2\hat{j}_{i}^{2}}   \left( \bar{n}_a  {n}_b  \bar{n}_c {n}_i  -  {n}_a  \bar{n}_b  {n}_c  \bar{n}_i \right) 
&\chi^\beta_{ij} =\! \sum_{abc J} \frac{\hat{J}^2}{2\hat{j}_{i}^{2}}   \left( \bar{n}_a  \bar{n}_b  {n}_c {n}_i  -  {n}_a  {n}_b  \bar{n}_c  \bar{n}_i \right) 
 \Omega^{J}_{ciab} \Gamma^{J}_{abcj} \\
%%%%%%%%%%%%%%%%%%%%%%%%%%%%%%%%%%%%%%%%%%%%%%%%
&\bar{\chi}^{\gamma ~ J}_{i \bar j k \bar l} = \!  \sum_{ab}\hat J^2 \left( {n}_a \bar{n}_b \bar{n}_k  {n}_l    -  \bar{n}_a  {n}_b {n}_k  \bar{n}_l  \right)   \bar{\Omega}^{J}_{i \bar j  a \bar b} \bar{\Omega}^{J}_{a \bar b  k \bar l} \\
%%%%%%%%%%%%%%%%%%%%%%%%%%%%%%%%%%%%%%%%%%%%%%%%
&\chi^{\delta~J}_{ijkl} =\!  \sum_{ab}\frac{\hat J^2}{4}\! \left(   {n}_a  {n}_b \bar{n}_k \bar{n}_l -    \bar{n}_a \bar{n}_b {n}_k {n}_l   \right) \Omega^{ J}_{ijab} \Omega^{ J}_{abkl}
\end{align}
\end{subequations}

The $J$-coupled factorized expressions for the two-body matrix elements of the double commutator are
 \begin{subequations}\allowdisplaybreaks[4]\label{Jcoupled_Factorized_DoubleCommutator_twobody}
\begin{align}
\Gamma^{\MyRomanNum{1} ~ J}_{ijkl}  =&  \sum_{a} \left\{ (1\! -\! \hat P^{J}_{ij} )   \chi^{\epsilon}_{ai} \Gamma^{J}_{ajkl} % \nonumber \\
%&
+ (1\! -\! \hat P^{J}_{kl} )   \chi^{\epsilon}_{ak} \Gamma^{J}_{ijal} \right\} \\
%%%%%%%%%%%%%%%%%%%%%%%%%%%%%%%%%%%%%%%%%%%%%%%%%%%%%%%%%%%%%%%%%
\Gamma^{\MyRomanNum{2} ~ J}_{ijkl}  =&  \sum_{a}
\left\{ (1 \!-\! \hat P^{J}_{ij} )    \chi^{\zeta}_{aj}  \Omega^{J}_{iakl} 
 %\nonumber \\
%&
-(1\! -\! \hat P^{J}_{kl} )   \chi^{\zeta}_{ak}  \Omega^{J}_{ijal} \right\}
\\
%%%%%%%%%%%%%%%%%%%%%%%%%%%%%%%%%%%%%%%%%%%%%%%%%%%%%%%%%%%%%%%%%%
\Gamma^{\MyRomanNum{3}_a J}_{ijkl}  =&   %\nonumber\\
%&\times
\sum_{ab} \left\{  (1 - \hat P^{J}_{ij} )  \chi^{\eta~J}_{ijab}  \Gamma^{J}_{abkl} \right. %\nonumber \\ 
%&
+  \left. (1 - \hat P^{J}_{kl} ) \Gamma^{J}_{ijab} \chi^{\eta~J}_{a b k l} \right\}\\
%%%%%%%%%%%%%%%%%%%%%%%%%%%%%%%%%%%%%%%%%%%%%%%%%%%%%%%%%%%%%%%%%%
\overline{\overline{ \Gamma}}^{\MyRomanNum{3}_bJ}_{j \bar l k \bar i} =& (1 - \hat P^{J}_{ij} )  (1 - \hat P^{J}_{kl} ) % \nonumber \\
%&\times
\sum_{ab} \overline{\overline{ \Gamma}}^{J}_{j \bar l a \bar b}  \left(   \overline{\overline{\chi}}^{\eta~J}_{ a \bar b k \bar i} 
  + \overline{\overline{ \chi}}^{\eta~J}_{k \bar i  b \bar a}  \right) \\ 
%%%%%%%%%%%%%%%%%%%%%%%%%%%%%%%%%%%%%%%%%%%%%%%%%%%%%%%%%%%%%%%%%%
\bar \Gamma^{\MyRomanNum{3}_c J}_{i \bar l k \bar j}  =& 
\frac{1}{2} (1 - \hat P^{J}_{ij} )  (1 - \hat P^{J}_{kl} ) \sum_{ab}  \bar \chi^{\theta~J}_{i\bar l a \bar b}  \bar \Gamma^{J}_{a \bar b k \bar j}
\\
%
%%%%%%%%%%%%%%%%%%%%%%%%%%%%%%%%%%%%%%%%%%%%%%%%%%%%%%%%%%%%%%%%%%
\Gamma^{\MyRomanNum{4}_a J}_{ijkl}  =&
-\sum_{ab} \left\{
(1 - \hat P^{J}_{ij} ) \chi^{\kappa~J}_{ijab} ~ \Omega^{J}_{abkl}    \right.  \nonumber \\
&\hspace{2em}+  \left. (1 - \hat P^{J}_{kl} ) \Omega^{J}_{ijab} ~ \chi^{\kappa~J}_{klab}  \right\} \\
%%%%%%%%%%%%%%%%%%%%%%%%%%%%%%%%%%%%%%%%%%%%%%%%%%%%%%%%%%%%%%%%%%
\overline{\overline{ \Gamma}}^{\MyRomanNum{4}_b J}_{j \bar l k \bar i} =& (1 - \hat P^J_{ij} )  (1 - \hat P^J_{kl} ) % \nonumber \\ 
%& \times
\sum_{ab} \overline{\overline{  \Omega}}^{J}_{j \bar l a \bar b} \left(   ~ \bar \chi^{\iota~J}_{a \bar b k \bar i}  
  -  \bar \chi^{\iota~J}_{ k \bar i a \bar b } \right) \\ 
%%%%%%%%%%%%%%%%%%%%%%%%%%%%%%%%%%%%%%%%%%%%%%%%%%%%%%%%%%%%%%%%%%
\bar \Gamma^{\MyRomanNum{4}_c~J}_{i \bar l k \bar j}  =&   \frac{1}{2} (1 - \hat P^J_{ij} )  (1 - \hat P^J_{kl} ) \sum_{ab} \bar \chi^{\lambda~J}_{i \bar l a \bar b}  ~ \bar\Omega^{J}_{a\bar k j \bar  b}  
\end{align}
\end{subequations}
and the $J$-coupled intermediates are
\begin{subequations}\allowdisplaybreaks[4]\label{chi_2b_Jcoupled}
\begin{align}
\chi^{\epsilon}_{ij} =& \frac{1}{2\hat{j}^2_j} \sum_{abc J} \hat{J}^2  \left( \bar{n}_a \bar{n}_b {n}_c    +  {n}_a  {n}_b \bar{n}_c \right)  \Omega^{J}_{ciab} \Omega^{J}_{abcj} \\  % I   ab
%%%%%%%%%%%%%%%%%%%%%%%%%%%%%%%%%%%%%%%%%%%%%%%%%%%%%%%%%%%%%%%%%%
\chi^{\zeta}_{ij}  =&\frac{1}{2} \sum_{abc J} \hat{J}^2 j^{-2}_j \left( \bar{n}_a  \bar{n}_b {n}_c  + {n}_a {n}_b \bar{n}_c  \right) \nonumber \\
 &\times  \Gamma^{J}_{aibc}  \Omega^{J}_{bcaj} \\   % IV ab
%%%%%%%%%%%%%%%%%%%%%%%%%%%%%%%%%%%%%%%%%%%%%%%%%%%%%%%%%%%%%%%%%%
\bar \chi^{\eta~J}_{i\bar jk \bar l} =& \sum_{ab} \left( \bar{n}_a  {n}_b  \bar{n}_k  +  {n}_a  \bar{n}_b  {n}_k \right) \bar \Omega^{J}_{i \bar j a \bar b} \bar \Omega^{J}_{a \bar b k \bar l} \\    % II  ac
%%%%%%%%%%%%%%%%%%%%%%%%%%%%%%%%%%%%%%%%%%%%%%%%%%%%%%%%%%%%%%%%%%%
\chi^{\theta~J}_{ijkl}  =&  \sum_{ab} \left(   {n}_a {n}_b \bar{n}_k  +  \bar{n}_a  \bar{n}_b {n}_k + {n}_a {n}_b \bar{n}_j  \right. \nonumber\\
&+  \left. \bar{n}_a  \bar{n}_b {n}_j \right)  \Omega^{J}_{ijab} \Omega^{J}_{abkl}  \\ 
% II  e f
%%%%%%%%%%%%%%%%%%%%%%%%%%%%%%%%%%%%%%%%%%%%%%%%%%%%%%%%%%%%%%%%%%%%
\bar \chi^{\iota~J}_{i\bar j k \bar l} =& \sum_{ab} \left(   \bar{n}_a {n}_b \bar{n}_k  +  {n}_a  \bar{n}_b  {n}_k \right) \bar\Gamma^{J}_{i \bar j a \bar b}   \bar \Omega^{J}_{a \bar b k \bar l} \\     % III ab
%%%%%%%%%%%%%%%%%%%%%%%%%%%%%%%%%%%%%%%%%%%%%%%%%%%%%%%%%%%%%%%%%%%%
\overline{\overline{  \chi}}^{\kappa~J}_{i \bar jk \bar l}  =&   \sum_{ab} \left(   \bar{n}_a {n}_b {n}_l  +  {n}_a  \bar{n}_b  \bar{n}_l \right) 
    \overline{\overline{  \Omega}}^{J}_{i \bar j b \bar a} \overline{\overline{  \Gamma}}^{J}_{b \bar a k \bar l} \\     % III cd
%%%%%%%%%%%%%%%%%%%%%%%%%%%%%%%%%%%%%%%%%%%%%%%%%%%%%%%%%%%%%%%%%%%%%
\chi^{\lambda~}_{ijkl}  =& \sum_{ab} \left(   \bar{n}_a \bar{n}_b {n}_l +  {n}_a  {n}_b  \bar{n}_l \right)  \Gamma^{J}_{ijab} \Omega^{J}_{abkl}    \nonumber\\   
 &+\sum_{ab} \left(   \bar{n}_a \bar{n}_b {n}_j +  {n}_a  {n}_b  \bar{n}_j \right)  \Omega^{J}_{ijab} \Gamma^{J}_{abkl}  
% III ef
\end{align}
\end{subequations}
where $\hat P^{J}_{ij} \equiv (-1)^{ j_i  + j_j - J  } P^{J}_{ij} $ is the J-coupled permutation operator.

We note that the intermediate operators appearing in 
equations~\eqref{Jcoupled_Factorized_DoubleCommutator_twobody} do not always have the same coupling scheme with which they are defined in equations~\eqref{chi_2b_Jcoupled}.
The translation between these can be obtained from equations~\eqref{Pandya} and \eqref{cross-coupled_Pandya}.

In practice, one can derive an equation to manage the direct transformation instead of taking two steps.

\bibliography{references}% Produces the bibliography via BibTeX.

\end{document}